\documentstyle[aps,preprint,epsf]{revtex}

\begin{document}

\draft

\title{A pseudopotential study of electron-hole excitations in colloidal,
free-standing InAs quantum dots}

\author{A.~J.~Williamson and Alex~Zunger}
\address{National Renewable Energy Laboratory, Golden, CO 80401}

\date{\today}
\maketitle

\begin{abstract}
Excitonic spectra are calculated for free-standing, surface passivated
InAs quantum dots using atomic pseudopotentials for the
single-particle states and screened Coulomb interactions for the
two-body terms.  We present an analysis of the single particle states
involved in each excitation in terms of their angular momenta and
Bloch-wave parentage.  We find that (i) in agreement with other
pseudopotential studies of CdSe and InP quantum dots, but in contrast
to k.p calculations, dot states wavefunction exhibit strong odd-even
angular momentum envelope function mixing (e.g. $s$ with $p$) and
large valence-conduction coupling. (ii) While the pseudopotential
approach produced very good agreement with experiment for
free-standing, colloidal CdSe and InP dots, and for self-assembled
(GaAs-embedded) InAs dots, here the predicted spectrum does {\em not}
agree well with the measured (ensemble average over dot sizes)
spectra.  (1) Our calculated excitonic gap is larger than the PL
measure one, and (2) while the spacing between the lowest excitons is
reproduced, the spacings between higher excitons is not fit well.
Discrepancy (1) could result from surface states emission.  As for
(2), agreement is improved when account is taken of the finite size
distribution in the experimental data.  (iii) We find that the single
particle gap scales as $R^{-1.01}$ (not $R^{-2}$), that the screened
(unscreened) electron-hole Coulomb interaction scales as $R^{-1.79}$
($R^{-0.7}$), and that the eccitonic gap sclaes as $R^{-0.9}$.  These
scaling laws are different from those expected from simple models.
\end{abstract}

\section{Introduction}

Semiconductor quantum\cite{mrs_feb98} dots have recently generated
considerable interest due to the dramatic differences between their
electronic structure and that of the bulk material from which they are
derived.  These differences arise from the lower symmetry of the dot,
quantum confinement level shifts and the enhancement of electron hole
Coulomb and exchange interactions.

While the Stranski-Krastanov growth technique produces
semiconductor-embedded dots\cite{samuelson94,petroff96,grundman95}
which are typically subject to only a small confining potential,
colloidal chemistry techniques\cite{bawendi96,banin96,micic96} produce
dots whose surfaces are passivated by large organic molecules.  This
leads to much larger confining potentials (a few eV), and thus many
more confined energy levels.  These colloidal techniques have recently
enabled the production of quantum dots with diameters from 10 to
60\AA~, made from CdSe\cite{bawendi96:2}, CdS\cite{colvin91},
InP\cite{micic98} and InAs\cite{alivisatos98}.  The spectroscopy of
these systems is richer than that of the SK dots, with
several groups reporting data for up to 10 excited states in
CdSe\cite{bawendi96:2}, InP\cite{micic98} and InAs\cite{alivisatos98}.
On the theoretical side, colloidal CdSe\cite{bawendi96:2,wang_cdse} and
InP\cite{micic98,fu98} have recently been studied using both the
standard 6x6 k.p method\cite{bawendi96:2,micic98} and more
sophisticated pseudopotential\cite{wang96,fu98,wang_cdse} techniques.

In this paper we turn our attention to colloidally grown InAs quantum
dots.  InAs is a challenging system to study because, (i) the small
bulk band gap of 0.42 eV suggests strong valence-conduction band
coupling.  This is supported by recent
studies\cite{alivisatos98,mizel98} showing that the standard 6x6 k.p
formalism fails to describe the observed electronic transitions, while
8x8 k.p produces a better fit to measured excitonic transitions, (ii)
the small band gap combined with a large spin-orbit splitting (0.38
eV) suggests that spin orbit effects will play a significant role,
(iii) the observed band gap extends to 1.6 eV, i.e. almost four times
the bulk value, suggesting dramatic quantum confinement, (iv) the
large confining potential leads to as many as eight clearly resolved
excitations, for a range of dot sizes.

We use a multi-band pseudopotential Hamiltonian to calculate both the
single particle states and electron-hole excitation energies for InAs
quantum dots with a range of sizes.  Our method is different from
another recent pseudopotential calculation of Mizel and
Cohen\cite{mizel98} in that although their formalism can be
generalized to multiple bands, they use only a single-band
approximation (like the EMA or the Truncated Crystal
Method\cite{zhang93}) and do not include electron-hole effects.  Both
methods however, do incorporate non-parabolicity of the bands.  The
nature of the single particle states is analyzed in terms of the
parent Bloch orbitals and envelope functions, which enables a detailed
description of the origin of each calculated exciton.  This analysis
allows comparisons to be made with existing effective mass based
predictions for the single particle states in these dots.  In contrast
to envelope function based effective mass calculations, our more
general treatment shows that the single particle states include a
significant amount of mixing of valence and conduction states, as well
as a mixing of envelope functions with odd and even parity.  We have
calculated the {\em single dot} absorption spectra of a few dots.
Comparing with the experimental results for ensembles of dots sizes,
we do not find good agreement in contrast to the k.p
method\cite{alivisatos98} that gives better agreement.  However,
agreement with our results is improved when the finite size
distribution of the experimental dot samples is taken into account.
We discuss possible reasons for the conflicts between experiment and
theory.

\section{Pseudopotential calculations of the electronic structure of
InAs quantum dots}\label{method}
Our pseudopotential calculations consist of two steps.  First we
calculate the single particle states for the quantum dot
(Sections.~\ref{single} and \ref{analysis}).  Then we
solve the ``two-particle'' problem by calculating the energy of the
electron-hole excitations of the system (Section.~\ref{spectra_sect}).

\subsection{Single Particle Calculation}\label{single}
We use a pseudopotential Hamiltonian to model the single-particle
electronic structure of the system.
\begin{equation}\label{hamiltonian}
\hat{H} = -\frac{1}{2}\nabla^2+\sum_{\alpha,n}v_\alpha({\bf r}-{\bf
R}_{\alpha n}) + v_\alpha^{(SO)} \;\;\;.
\end{equation}
The system's potential is constructed from a sum of screened atomic
pseudopotentials, $v_\alpha$, where $\alpha$ represents In and As, and
${\bf R}_{\alpha n}$ are the positions of the In and As atoms within
the dot.  The pseudopotentials, $v_\alpha$, have been
fitted to the experimental band gaps and effective
masses of bulk InAs.  The experimental and fitted values are given in
Table~\ref{fit_results}.  The fitted InAs bulk band structure is
shown in Fig.~\ref{bulk}.  We use the reciprocal space functional form
of the pseudopotential
\begin{equation}\label{epm}
v_\alpha({\bf q}) = a_{0\alpha} {(q^2-a_{1\alpha})\over 
a_{2\alpha}e^{a_{3\alpha}q^2}-1} \;\;\; ,
\end{equation}
where $a_{0\alpha},a_{1\alpha},a_{2\alpha},a_{3\alpha}$ are adjustable
parameters.  The fitted values of $a_{i,\alpha}$ are given in
Table~\ref{values_table}.  The same pseudopotential form was recently
used to study self-assembled GaAs-covered InAs pyramidal
dots\cite{williamson98:2,jkim98,williamson98:3,wang98}.  In
Refs.\onlinecite{williamson98:2,jkim98,williamson98:3,wang98} the
pseudopotential contained an additional term to describe the effects
of strain in the system.  As the free-standing InAs dots studied here
are strain free this term is not required here.  We assume the bulk
zinc-blende structure and the bulk In-As inter-atomic spacing
(d=6.057\AA).  We construct spherical InAs dots of radius $R$, by
selecting all the atoms that fall within a sphere of this radius. Any
atoms from the surface of the dot which have three dangling bonds are
removed.  The result of adopting this atomistic description is to
reduce the symmetry of the dots from the full spherical symmetry of
continuum models to the lower T$_d$ symmetry.  The sizes and
compositions of the four dots studied in this paper are listed in
Table~\ref{dots}.

To simulate the chemical passivation via ligands\cite{micic96}, we
embed the InAs dots within an artificial barrier material, represented
by an atomic pseudopotential, fitted to have a larger band gap than
InAs, thus producing a type-I alignment between the dot and the
barrier.  The barrier material is designed to have the InAs lattice
constant so that no strain is introduced into the system.  This
embedding process is physically equivalent to the choice of a finite
barrier surrounding the dots and is designed to reflect the fact that
the dots are surrounded by organic molecules which themselves have
large, but finite band gaps\cite{note1}.  

We expand the single particle wavefunctions, $\psi_i$, in a plane wave
basis whose cutoff must be identical to that used in the original
pseudopotential generating procedure,
\begin{equation}\label{basis}
\psi_i({\bf r})=\sum_{{\bf G}}^{E_{cut}} c_{{\bf G},i} e^{i{\bf
G}.{\bf r}} \;\;\; .
\end{equation}
The matrix elements of the Hamiltonian in Eq.(\ref{hamiltonian}) in
the basis of Eq.(\ref{basis}) are calculated according to
\begin{equation}
\hat{H}_{{\bf G},{\bf G}'} = \frac{1}{2}{\bf G}^2\delta_{{\bf G},{\bf
G}'} + V_{local}({\bf G}-{\bf G}') + V_{nonlocal}({\bf G},{\bf G}')
\;\;\; .
\end{equation}
The spin orbit interaction is represented by a non-local
pseudopotential, $V_{nonlocal}({\bf G},{\bf G}')$, which is evaluated
in real space using the linearly scaling small box method from
Ref.\onlinecite{wang95}.  This method applies the non-local
pseudopotential to each atom in turn.  For each atom, a new
wavefunction $\psi_{box}({\bf r})=\psi_i({\bf r})$ is defined within a
small box around the atom. $\psi_{box}({\bf r})$ is periodically
repeated and then fast Fourier transforms are used to generate
$\psi_{box}({\bf G})$.  The non-local pseudopotential is then applied
by
\begin{equation}
\phi_i({\bf G}) = \sum_{{\bf G}'}V_{nonlocal}({\bf G},{\bf
G}')\;\;\psi_{box}({\bf G}') \;\;\; .
\end{equation}

The InAs dots, surrounded by barrier material, form a supercell which
is periodically repeated.  Sufficient barrier atoms are used, to
ensure that the interactions between an InAs dot and its periodic
images are negligible.  The total number of atoms (In, As and barrier)
in the largest supercell is 25,000 atoms, which is too large for the
Hamiltonian in Eq.(\ref{hamiltonian}) to be solved by conventional
diagonalization methods.  We thus use the ``Folded
Spectrum Method'' (FSM) \cite{wang94,kamat}, in which one solves for
the eigenstates of the equation
\begin{equation}\label{fsm}
\left( \hat{H}-\epsilon_{ref} \right)^2\psi_i
= (\epsilon-\epsilon_{ref})^2\psi_i \;\;\; ,
\end{equation}
where $\epsilon_{ref}$ is a reference energy.  By placing
$\epsilon_{ref}$ within the band gap of the dot, and close to the
valence band maximum or conduction band minimum, one is then able to
obtain the top few valence states or the bottom few conduction states
respectively by minimizing $<\psi|(H-E_{ref})^2|\psi>$.  The use of
FFTs results in an $N\ln N$ scaling with the number of atoms, $N$,
enabling calculation of systems containing 10$^5$ atoms.  It is easy
to guess an approximate value of $\epsilon_{ref}$: We first determine
the position of the {\em bulk} InAs band edges by using the same
pseudopotential to calculate the bulk InAs band structure,
$\epsilon_{nk}$. This is illustrated in Fig.\ref{bulk}.  As quantum
confinement effects push electron(hole) levels up(down) in energy,
placing $\epsilon_{ref}$ within the bulk band gap ensures that it will
also be within the band gap of the InAs dot.  Note, that although
Eq.(\ref{fsm}) has twice as many eigensolutions as the standard
Schr\"{o}dinger equation, $\hat{H}\psi_i = \epsilon_i\;\psi_i$, we
only search for solutions to Eq.(\ref{fsm}) in a finite basis of $N$
plane waves [Eq.(\ref{basis})].  We therefore only have $N$ solutions
to Eq.(\ref{fsm}).  By applying the $\hat{H}$ operator twice, it can
be shown that each of these $N$ solutions is a solution of
$\hat{H}\psi_i = \epsilon_i\;\psi_i$ and thus all our solutions to
Eq.(\ref{fsm}) are also solutions of the standard Schr\"{o}dinger
equation.  Note also that our approach to the single-particle problem
includes [Eq.({\ref{hamiltonian})] shape effects, interface effects
and spin-orbit coupling in the Hamiltonian [Eq.({\ref{hamiltonian})].
In the effective mass approximation, on the other hand, these effects
are included perturbatively\cite{alivisatos98}.  Also note that we
neglect self-consistency with respect to the bulk.  As the dots
contain several hundred atoms we expect the potential of the dot
interior to be bulk-like.  The pseudopotential used here was fitted to
reproduce these bulk properties without requiring a self-consistent
calculation.

\subsection{Methods for analyzing the single particle wavefunctions}\label{analysis}

Having calculated the single particle states of an InAs quantum dot,
we are interested in analyzing the nature of each of these states in
terms of the bulk Bloch wavefunctions $\{\phi_{nk}\}$ from which they
are derived, the symmetry of their envelope functions and their
defining quantum numbers.  These quantities will enable us to classify
each of the electron-hole excitations calculated in the following
section in terms of the detailed nature of their initial valence state
and the final conduction state.  They also enable us to make contact
with alternative electronic structure theories, such as the k.p
technique, which typically classifies states in terms of their
envelope function and total angular momentum.  Note, the expressions
in Eqs.(\ref{eq7})-(\ref{weight}) below are used only to analyse the
solutions of Eq.(\ref{basis}) and do not affect the results.

We first deconstruct each single particle dot wavefunction,
$\psi_i({\bf r})$, in terms of the bulk Bloch states at the $\Gamma$
point, $\phi_{n\Gamma}=u_{n\Gamma}({\bf r})e^{i{\bf k}.{\bf r}}$ in
the manner described in Ref.\onlinecite{wang_cdse}.  We choose this
basis to (i) enable the calculation of the different angular momentum
character of the states, and (ii) to enable comparison with
conventional envelope function based methods.  These zone center Bloch
states form a complete set, and therefore one can expand the single
particle dot wavefunction according to
\begin{equation}\label{eq7}
\psi_i({\bf r}) = \sum_{n}\left[ \sum_{{\bf k}}b_n({\bf k})e^{i{\bf
k}.{\bf r}}\right]u_{n,\Gamma}({\bf r})
\end{equation}
which can be rewritten as
\begin{equation}\label{decomp}
\psi_i({\bf r}) = \sum_n u_{n,\Gamma}({\bf r})f_n^{(i)}({\bf r}) \;\;\; ,
\end{equation}
where $u_{n,\Gamma}({\bf r})$ is the bulk Bloch wavefunction for the
$n$th band at the $\Gamma$ point, and $f_n^{(i)}({\bf r})$ is a
corresponding envelope function.  Note that the ``single band
approximation'' corresponds to retaining a single $n$ value in
Eq.(\ref{decomp}), while the 6x6 k.p corresponds to $n=$ bulk VBM
(sixfold degenerate, including spin).  The present ``multiband'' method
corresponds to a large number of bands in Eq.(\ref{decomp}).  We
further decompose each envelope function, $f_n^{(i)}({\bf r})$, into
spherical harmonics, $Y_{L,m}(\theta,\phi)$, according to,
\begin{equation}
f_n^{(i)}({\bf r})=\sum_{L,m}R_{n,L,m}^{(i)}(|{\bf r}|)Y_{L,m}(\theta,\phi) \;\;\;,
\end{equation}
and then define the weight from each angular momentum component,
$w_{n,L}^{(i)}$, as
\begin{equation}\label{weight}
w_{n,L}^{(i)}=\sum_{m=-L}^{L}\int_0^{R}|rR^n_{L,m}(r)|^2dr \;\;\;, 
\end{equation}
where $R$ is the radius of the quantum dot.  The quantity,
$w_{n,L}^{(i)}$, tells us how much of bulk band $n$ and angular
momentum $L$ is contained in the dot wavefunction $\psi_i$.

An additional property that is useful for classifying the single
particle states is the total angular momentum, $F$, of dot state $i$.
This is calculated for each of the single particle dot states
according to
\begin{equation}
F(F+1)=\left< \sum_n
u_{n,\Gamma}f_n^{(i)}|(\hat{F}_u+\hat{F}_f)^2|\sum_m u_{m,\Gamma}
f_m^{(i)}\right> \;\;\; .
\end{equation}
The angular momentum operator $\hat{F}_u = i{\bf r}\times\nabla
+\vec{s}$, (where $\vec{s}$ is the Dirac spin matrix), acts only on the
Bloch orbitals, $u_n({\bf r})$, and $\hat{F}_f = i{\bf r}\times\nabla$
acts only on the envelope function, $f_n^{(i)}({\bf r})$.  

\subsection{Two particle calculation and the absorption spectrum}\label{spectra_sect} 

Calculation of an optical absorption spectrum based on the single
particle states of a quantum dot, requires a solution for the
``two-body'' electron-hole problem.  More complete solutions of the
electron-hole problem via the configuration interaction require
knowledge of the Coulomb, exchange and correlation energy associated
with the electron-hole pair\cite{franceschetti99}.  In this work, we
choose to ignore the fine structure effects resulting from the
exchange and correlation energy, reducing the two-body problem simply
to calculating the electron-hole Coulomb energy.  We then define the
exciton energy in terms of the initial single particle valence state
energy, $\epsilon_i$, the final conduction state energy, $\epsilon_j$,
and the electron-hole Coulomb energy, $J_{ij}$, as
\begin{equation}\label{exciton_energy}
E_{exciton}^{(ij)}=(\epsilon_j-\epsilon_i)-J_{ij} \;\;\; .
\end{equation}

The electron-hole Coulomb energy, $J_{ij}$, between each of
the possible electron(i) - hole(j) pairs is calculated by a direct
integration of the single particle wavefunctions
\begin{equation}\label{Coulomb_energy}
J_{ij}(R)=\int\int\frac{|\psi_i({\bf r}_1)|^2|\psi_j({\bf
r}_2)|^2}{\overline{\epsilon}({\bf r}_1-{\bf r}_2;R)|{\bf r}_1-{\bf
r}_2|} d{\bf r}_1 d{\bf r}_2 \;\;\; \propto R^{-n},
\end{equation}
where $\psi_i({\bf r})$ and $\psi_j({\bf r})$ are our calculated
electron and hole wavefunctions and $\overline{\epsilon}({\bf
r}_1-{\bf r}_2,R)$ is a distance dependent screening function, connected
to the dielectric function, $\epsilon({\bf r},{\bf r}',R)$, by,
\begin{equation}\label{screen1}
\frac{1}{\overline{\epsilon}({\bf r},{\bf r}'')\;|{\bf r}-{\bf r}''|}
=\int d{\bf r}'\;\;\epsilon^{-1}({\bf r},{\bf r}';R)\frac{1}{|{\bf
r''}-{\bf r}'|}
\end{equation}
and $n$ is a size scaling exponent discussed below.  We follow
Ref.\onlinecite{wang_cdse} in writing the Fourier transform of
$\epsilon^{-1}({\bf r}_1-{\bf r}_2;R)$ as the sum of electronic and
ionic terms,
\begin{equation}
\epsilon^{-1}(k;R)=\epsilon_{el}^{-1}(k;R)+\Delta\epsilon_{ion}^{-1}(k;R)
\;\;\; ,
\end{equation}
The electronic term is defined in terms of the Thomas Fermi model of
Resta\cite{resta77},
\begin{equation}\label{e_ion}
\epsilon_{el}^{-1}(k;R)=\frac{k^2+q^2\sin(kR_\infty)/
(\epsilon_{\infty}^{dot}kR_\infty)}{k^2+q^2}
\;\;\; .
\end{equation}\label{screen2}
Note, the $R$ dependence of $\epsilon_{el}^{-1}(k;R)$ comes from
$\epsilon_{\infty}^{dot}$. We define $q$ in Eq.(\ref{e_ion}) as
\begin{equation}
q=2\pi^{-1/2}(3\pi^2n_0)^{1/3} \;\;\; ,
\end{equation}
where $n_0$ is the electron density, and $R_\infty$ is the solution to
\begin{equation}
\frac{\sinh(qR_\infty)}{qR_\infty} = \epsilon_\infty^{dot} \;\;\; .
\end{equation}
The ionic term is taken from Ref.\onlinecite{haken}
\begin{equation}
\Delta\epsilon_{ion}^{-1}(k;R)=\left(\frac{1}{\epsilon_0^{dot}}-\frac{1}{\epsilon_\infty^{dot}}\right)\left(\frac{1/2}{1+\rho_h^2k^2}+\frac{1/2}{1+\rho_e^2k^2}\right)
\;\;\; .
\end{equation}
Here $\rho_{h,e}=(2m_{h,e}^*\omega_{LO}/\hbar)^{-1/2}$, $\omega_{LO}$
is the longitudinal optical-phonon frequency (238.6 cm$^{-1}$) and
$m_{h,e}^*$ are the electron and hole effective masses (0.023 and 0.4  
$m_0$).  
To obtain the value of $\epsilon_0^{dot}$, we again assume that the
dot interior is bulk-like so $\epsilon_0^{dot}-\epsilon_\infty^{dot}$
can be approximated from the bulk,
\begin{equation}
\epsilon_0^{dot}-\epsilon_\infty^{dot}=\epsilon_0^{bulk}-\epsilon_\infty^{bulk}=\Delta\epsilon_{ion}^{bulk}
\;\;\; .
\end{equation}
We use the bulk value of 2.9 for $\Delta\epsilon_{ion}^{bulk}$.
Finally, we use a generalization\cite{tsu} of the Penn
model\cite{penn} to obtain $\epsilon_\infty^{dot}$,
\begin{equation}\label{eq21}
\epsilon_\infty^{dot}(R)=1+(\epsilon_\infty^{bulk}-1)\frac{(E_{gap}^{bulk}+\Delta)^2}{(E_{gap}^{dot}(R)+\Delta)^2} 
\;\;\; ,
\end{equation}
where $\Delta=E_2-E_{gap}^{bulk}=4.28$ eV.  The electronic and ionic
components of the screening function are plotted in
Figs.~\ref{dielectric_recip}(a) and (b).  One sees that the electronic
contribution dominates for all but the smallest
$k$-vectors. Figure~\ref{dielectric_real} shows the screened
dielectric function in real space, $\epsilon (R,r)$, as a function of
electron-hole separation $r$, for dots with radii, $R$, from
Table~\ref{dots}.  At very small separations, $\epsilon (R,r)$
approaches 1.0, which corresponds to no screening at zero separation
between the electron and hole.  When the electron and hole are
separated by $r=6-8$\AA, $\epsilon (r)$ approaches the particular
value of $\epsilon_\infty^{dot}(R)$ which is different for each of the
dots : The resulting values of $\epsilon_\infty^{dot}(R)$ are 7.19,
7.91, 8.41 and 8.92 for dot diameters 23.9, 30.3, 36.6 and 42.2\AA.
The ionic screening, $\Delta\epsilon_{ion}$ has a very long tail, such
that $\epsilon (r)$ approaches $\epsilon_0^{dot}$ for $r>100$\AA.  The
R-dependence of $\epsilon (r,R)$ can be clearly seen in
Fig.~\ref{dielectric_real}.  In all cases the value is smaller than
the bulk InAs value of $\epsilon_0^{bulk}=15.15$.

More simplified calculations of $J_{ij}$ assume a universal value for
all $i$ and $j$, based on an $s$-like envelope function and
an infinite potential barrier\cite{brus86}.  These approximations lead
to the often adopted formula
\begin{equation}\label{brus_coulomb}
J^{EMA}=\frac{-3.572}{\epsilon_0^{bulk}2R} \;\;\; \propto R^{-1} .
\end{equation}
Our pseudopotential calculated values for the screened electron-hole
Coulomb energy between an electron in the lowest energy conduction
state and a hole in the highest energy valence state, $J_{11}$, are
plotted along with those predicted by Eq.(\ref{brus_coulomb}) in
Fig.~\ref{coulomb_fig}.  The magnitude of the pseudopotential values
is larger than that predicted by Eq.(\ref{brus_coulomb}) for all sizes
of dot.  A fit of the dependence of the screened Coulomb energies,
$J$, on the dot radius, $R$, to $J=A/R^n$, yields a size exponent,
$n$, of 1.79, for the pseudopotential results, compared to the 1.0
scaling predicted by the Eq.(\ref{brus_coulomb}).  The difference in
magnitude between Eq.(\ref{brus_coulomb}) and pseudopotential
calculated Coulomb energies of Eq.(\ref{Coulomb_energy}) has two
sources: Firstly, we use microscopic (rather than envelope) functions
that are not required to vanish exactly on the surface of the dot as
EMA functions are required to do when subjected to an infinite
potential barrier.  This wavefunction effect causes the unscreened
Coulomb energy $\epsilon J$ to scale as $R^{-m}$ with $m<1$, while in
the EMA $m=1$.  This effect was discussed by Franceschetti {\em
et. al.}\cite{franceschetti97} who found $m\sim 0.7$ for GaAs.
Secondly, the description of screening in Eqs.(\ref{screen1}) to
(\ref{eq21}) leads to a reduced screening at small $R$ with respect
the bulk screening, $\epsilon_0^{bulk}$.  This acts to {\em increase}
our values of $J_{ij}$ and its $R$ dependence with respect to those in
Eq.(\ref{brus_coulomb}).  This effect is oppostie and stronger than
the wavefunction effect.  The net effect is $J\sim R^{-1.79}$ compared
with $J\sim R^{-1}$ in Eq.(\ref{brus_coulomb}).  The present
calculated values for the size scaling of the unscreened electron-hole
Coulomb interaction, $\epsilon J$, and the screened Coulomb
interaction, $J$ for InAs dots are compared with calculations for
Si\cite{reboredo99}, InP\cite{fu97} and CdSe\cite{wang_cdse} dots as
well as the predictions from Eq.(\ref{brus_coulomb}) in
Table~\ref{scalings}.  It shows that for the scaling of $\epsilon J$
has $m<1$ while the scaling of $J$ has $n>1$.

The final stage in the calculation of our theoretical absorption
spectra, is to calculate the dipole matrix transition element for the
transition from valence state $i$ to conduction state $j$,
$M_{i\rightarrow j}$,
\begin{equation}
M_{i\rightarrow j}(R) = \left| \left< \psi_i|\nabla|\psi_j\right>
\right|^2
\;\;\; .
\end{equation}
The optical absorption spectra, $I(E,R)$, can then be written as,
\begin{equation}\label{spectra}\label{ensemble_eq}
I(E,R)=\sum_i\sum_j M_{i\rightarrow j}(R)
\exp\left[-\left(\frac{E-E_{exciton}^{(ij)}}{\sigma}\right)^2\right]
\;\;\; ,
\end{equation}
where $E_{exciton}^{(ij)}$ is the excitation energy
[Eq.(\ref{exciton_energy})] and $\sigma$ represents an experimental
linewidth broadening which we set to 25 meV.

Our theory of Eq.(\ref{spectra}) corresponds to the spectra of a
single dot with a well defined radius, $R$, and shape.  Current
samples exhibit a size distribution about an average size,
$\overline{R}$, producing an ensemble absorption spectra
\begin{equation}
\overline{I}(E,<R>) = \int P(R) I(E,R) dR
\end{equation}
where $P(R)$ is the size distribution function.  A major contribution
in the interpretation of experimental data is that if the size
distribution is too broad, a given transition $(ij)$ for size $R_1$
could overlap with another transition (i'j') for size $R_2$.

\section{Results}\label{results}
\subsection{Single Particle States}\label{single_results}
Using the Hamiltonian described in Section\ref{single}, the 48 highest
energy hole states and the 24 lowest energy electron states were
calculated for the InAs dots listed in Table~\ref{dots}.
Figure~\ref{contour} shows plots of the electron and hole
wavefunctions squared, for the dot with a diameter of 42.2\AA, in a
plane through the center of the dots.  The plots show that as a result
of the large band offsets between the vacuum and the InAs dot, the
wavefunctions are strongly localized within the InAs dot.
Approximately 90.5\% and 95.9\% of the ground state electron and hole
wave function squared respectively, are localized within the dot.

The single-particle energies of the highest energy hole state and the
lowest energy electron state are plotted as a function of diameter in
Fig.~\ref{vbm-cbm}.  For comparison, the same states obtained from a
recent Wannier function based, {\em single-band} pseudopotential
calculation of Mizel and Cohen\cite{mizel98} are also shown.
Figure~\ref{vbm-cbm} shows that in both calculations, as the size of
the InAs dot increases, the quantum confinement decreases, such that
the VBM and CBM approach the bulk VBM and CBM values.  We have fitted
the size dependence of the dot VBM and CBM according to
\begin{eqnarray}\label{scaling}
\epsilon_{vbm}^{dot} & = & E_{vbm}^{bulk} +
\frac{a}{R^{n_v}} \nonumber \\ 
\epsilon_{cbm}^{dot} & = & E_{cbm}^{bulk} + 
\frac{b}{R^{n_c}} \;\;\; ,
\end{eqnarray}
where $R$ is the radius of the InAs dot.  This fitting reveals that
the size exponents, $n_{v,c}$, of the hole and electron quantum
confinements are significantly different to the effective mass exponents:
\begin{equation}
n_v = \left\{ \begin{array}{ll}
		1.21 & \mbox{Current Pseudopotential} \\
		1.59 & \mbox{Single band}\cite{mizel98} \\
		2.0  & \mbox{EMA}
		\end{array} \right. 
n_c = \left\{ \begin{array}{ll}
		0.95 & \mbox{Current Pseudopotential} \\
		1.23 & \mbox{Single band}\cite{mizel98} \\
		2.0  & \mbox{EMA}
		\end{array} \right. \;\;\; .
\end{equation}
Figure~\ref{vbm-cbm} also shows that while the CBM states calculated
with the multi-band pseudopotential and the single-band method are
almost identical in energy, the VBM states calculated using the
single-band method are higher in energy than those calculated using
the pseudopotential.  This difference can be attributed to the fact
that the real dot valence band has strong interband coupling, so the
single band approximation used by Mizel and Cohen\cite{mizel98} is
more severe for the valence band.  For the CBM states, the single band
approximation is sufficient because the energy spacing between bands
is large and there is therefore little interband coupling.

{\em Single Particle Gaps:} In Fig.~\ref{gaps}(a) we plot the single
particle band gaps obtained from our multi-band pseudopotential
calculations.  For comparison we also show the single particle gaps
obtained from the single band calculations\cite{mizel98} of Mizel and
Cohen and the 8 band k.p calculations from Ref\cite{banin98}.
Figure~\ref{gaps}(a) shows that the single band
calculations\cite{mizel98} tend to underestimate the single particle
gap with respect to the more converged calculations, as a result of
the underestimate of the valence band quantum confinement in single
band models.  The 8 band k.p and current multi-band pseudopotential
results show good agreement with each other.  In Fig.~\ref{gaps}(b) we
compare the multi-band pseudopotential calculations with a recent PL
experiment, by extracting the single particle gaps from experimental
results from Ref.\cite{banin96} by subtracting an approximate value
for the electron-hole Coulomb energy calculated from
$E_{coul}=3.572/2\epsilon R$ (Ref.\cite{brus86}).  Here $\epsilon$ is
the static dielectric constant of bulk InAs and $R$ is the dot radius
in atomic units.  We also show the results of the single-particle gap
extracted from a new STM tunnelling experiment\cite{banin99} on the
same colloidal InAs dots.  The multi-band pseudopotential calculations
predict larger gaps than those measured by PL.  This can have three
reasons (i) the observed emission could involve surface states below
the CBM and above the VBM and would therefore have a lower energy than
the calculated band to band values.  (ii) The calculation assumes a
single dot, while in the experiment there is an ensemble of dots.
Thus, the observed emission could be of low energy since the largest
dots in the ensemble, having the lowest CBM and highest VBM energies
emit.  In addition (3) measured sizes suffer from the notorious
difficulty of measuring size accurately.  If the measured TEM size is
underestimated, (especially for the samll dots) this could explain the
discrepancy.  The new tunellling STM experiments on similar InAs dots,
Banin {\em et. al.}  measure a single particle gap that is higher than
that extracted from the optical experiment and is in much better
agreement with the calculations.

We fit the single particle (sp) band gap to the following power law,
\begin{equation}\label{gap}
E_{sp}^{dot}=E^{bulk}_{sp}+AD^n \;\;\; ,
\end{equation}
taking the bulk gap as 0.42 eV.  Our pseudopotential results yield a
size exponent $n_{sp}=-1.01$.  This is considerably smaller than the
effective mass predicted value of -2.  In the previous section we have
seen that the Coulomb interaction $J_{ij}$ scales as $R^{-1.8}$, while
here we see that the single particle gap scales as $R^{-1.01}$.  Thus
the {\em relative} importance of Coulomb effects increases for small
dots.  This is the opposite of the effect predicted by single-band
effective mass predictions of $J(R)\sim R^{-1}$ and $E_{gap}\sim
R^{-2}$.  The scaling, $n$, of the single particle band gap is
compared with similar calculations for InP, CdSe and Si dots in
Table~\ref{scalings}.  

{\em Excitonic gaps:} In Fig.~\ref{gaps}(c) we plot the actual PL
results along with our multi-band pseudopotential excitonic results
calculated from Eq.(\ref{exciton_energy}) including the electron-hole
Coulomb energy screened according to Eq.(\ref{Coulomb_energy}).  The
single-particle pseudopotential results from Fig.~\ref{gaps}(a) are
shown for comparison.  The size exponent for the excitonic gap is
$n_{exciton}=-0.90$, smaller than that for the single particle gap
($n_{sp}=-1.01$).  Using the Coulomb energy from
Eq.(\ref{Coulomb_energy}) rather than Eq.(\ref{brus_coulomb}) improves
the agreement with experiment in Fig.~\ref{gaps}(b) by a factor of 3.
However, the fit is still not particularly good.  The reasons for this
were discussed above.

{\em Confinement energies:} We have also calculated the ratio of the
conduction band quantum confinement, $\Delta E_{CB}(R)$, to the
valence band quantum confinement, $\Delta E_{VB}(R)$, defined as
\begin{eqnarray}
\Delta E_{CB}(R) & = & \epsilon_{cbm}^{dot}-\epsilon_{cbm}^{bulk}
\nonumber \\
\Delta E_{VB}(R) & = & \epsilon_{vbm}^{dot}-\epsilon_{vbm}^{bulk}
\;\;\; .
\end{eqnarray}
Define $\xi(R)$ as
\begin{equation}
\xi(R) = \frac{\Delta E_{CB}(R)}{\Delta E_{VB}(R)}
\end{equation}
Our multi-band calculations for InAs dots find a value of $\xi(R)$
that decreases with $R$.  The best linear fit (constant $\xi(R)$) to
the data yields a ratio, $\xi(R)$, of 3.13 for the multiband
calculations.  For Si dots, $\xi(R)$ was recently
measured\cite{van_buuren} to have a value of 0.5.

Within the single-band effective mass approximation, assuming an
infinite potential barrier at the surface, the quantum confinement of
electrons and holes can be written as $\Delta\epsilon_{e,h} =
\hbar^2\pi^2/2m^*_{e,h}R^2$, where $m^*_{e,h}$ is the effective mass
of electrons and holes at the $\Gamma$ point and $R$ is the size of
the quantum dot.  The ratio of electron confinement to hole
confinement
\begin{equation}\label{qc_ema_ratio}
\xi(R) = \frac{\Delta E_{cbm}(R)}{\Delta E_{vbm}(R)} =
\frac{m_h^*}{m_e^*} = \mbox{const.}
\end{equation}
is therefore given by the ratio of the electron and hole effective
masses which is approximately 15.4.  This is significantly different
to our multiband calculated value of 3.13.  This difference can be
attributed to the fact that the pseudopotential calculated hole states
are derived form a mixture of the bulk heavy and light hole states.
					 
\subsection{The nature of the single particle wavefunctions}

The results of the single particle wavefunction analysis from
Eqs.(\ref{decomp})-(\ref{weight}) is given for the 42.2\AA~diameter
dot in Table~\ref{analysis_table_12}.  Details of the analysis for the
other dot sizes are given at Ref.\cite{web_site}.  Only the states
which contribute to the excitonic peaks discussed in
Section~\ref{spectra_sect} are listed in the table.  For each state,
the fraction of the wavefunction derived from the bulk $\Gamma$-like
split off, heavy ($\Gamma_{8v}$) and light hole ($\Gamma_{7v}$) and
lowest conduction band ($\Gamma_{6c}$) states is given (see
Fig.~\ref{bulk}).  For each of these bulk states, the fraction of the
total wavefunction derived from envelope functions with $s$, $p$ and
$d$ symmetry is also given.  The significant contributions to each
state are marked in bold.  Contributions less than 0.01 have been set
to zero.  For example, the highest energy hole state has a fraction of
$0.0+0.02+0.02=0.04$ derived the bulk split-off band,
$0.50+0.29+0.09=0.89$ derived from the bulk heavy and light hole bands
and $0.0+0.02+0.0=0.02$ derived from the lowest bulk conduction band.
The remaining 0.05 fraction is derived from bulk bands further from
the band gap, and from higher angular momentum envelope functions.

Analysis of the results in Table~\ref{analysis_table_12} and those for
the other size dots reveals several interesting properties of the
single particle wavefunctions:

(1) The origin of the lowest lying electron states in the dot
follows qualitatively the predictions of single band effective mass
theory.  For example, the lowest electron state of the
42.2\AA~diameter dot (see Table~\ref{analysis_table_12}) is 69\%
derived from the bulk conduction band edge state ($\Gamma_{6c}$) with
an $s$-like envelope function.  The next two highest electron levels
are 64\% derived from the same bulk Bloch state, but with a $p$-like
envelope functions.

(2) As a result of the small band gap of bulk InAs, there is a
strong coupling between the electron and hole states.  Approximately
20\% of the weight of the lowest energy {\em electron states} in the
dot is derived from the split-off, heavy and light hole
($\Gamma_{8v}+\Gamma_{7v}$)) bulk states.  This valence-conduction
mixing explains why the 6x6 k.p method, which ignores such coupling,
fails to describe these states in InAs dots.

(3) The highest energy hole states in the dot have significant weight
from both the $s$ and $p$ envelope functions from both heavy and light
hole bulk states.  They therefore cannot be described, even
qualitatively, by a single band model.  The VBM of the
42.2\AA~diameter dot (see Table~\ref{analysis_table_12}) has 50\% $s$
and 29\% $p$ character.  Similarly, the lowest electron state in the
dot has 27\% non-$s$ character, originating from valence bands.  Such
an $s$-$p$ mixing is largely absent in current theoretical
descriptions of InAs dots via the k.p method\cite{alivisatos98}.

(4) The order of the electron and hole states changes with size.  This
reflects different size scaling of the quantum confinement for
different states.  For example in the 42.2\AA~diameter dot (see
Table~\ref{analysis_table_12}) the lowest conduction state is $s$-like
(69\%), then states 2,3 and 4 are $p$-like (64\%), then states 5,6,7
and 8 (not shown in the table) are a mixture of $d$ and $g$.  In the
30.3\AA~diameter dot, the states with 42\% $p$ character have moved
above the states identified by their significant $d$ and $g$ character
making them states 6,7 and 8 as opposed to 2,3 and 4.  Similarly, in
the valence band the state with 37\% $s$-like contribution from the
split off band and 13\% and 12\% from $p$ and $d$ envelopes of heavy
and light hole states is the 5$^{th}$ and 6$^{th}$ state from the top
of the valence band in the 30.3\AA~diameter dot but moves to the
11$^{th}$ and 12$^{th}$ state from the top of the valence band in the
42.2\AA~diameter dot.

\subsection{Calculated Single-dot absorption spectra}\label{spectra_results}
Theoretical single-dot absorption spectra were calculated for each of
the dots listed in Table~\ref{dots} using the method described in
section~\ref{spectra_sect}.  These spectra are shown in
Figs.~\ref{spectra_fig}(a)-(d).  The identities of each of the major
peaks in the spectra were determined by examining the nature of the
initial and final single particle states contributing to each peak.
The following criteria were used to establish these identities.
\begin{enumerate}
\item The ratio of the contributions to the single particle states from
the heavy hole, light hole, split off and conduction states.
\item The contribution of each angular momentum component from the
heavy hole, light hole, split off and conduction states.
\item The total angular momentum, F, of each single particle state
(see section~\ref{analysis}).
\item The strength of the dipole transition probability matrix element
for each peak.
\end{enumerate}
Table~\ref{peaks_table_12} shows the identities of the peaks for the
42.2\AA~diameter dot.  Details of the analysis for the other dot sizes
are given at Ref.\cite{web_site}.  They show that there are two
separate manifolds of transitions: peaks (a) to (e) which correspond
to transitions from the dot hole states to the {\em lowest} $s$-like
electron state and peaks (f) to (k) which correspond to transitions
from the dot hole states to the {\em next highest} $p$-like electron
state.  The analysis of the peaks shows:

(a) ``peak a'' represents the fundamental band gap transition.  The
initial valence state associated with this peak is a doubly degenerate
hole state with $s$ and $p$ character derived from both heavy and
light hole bulk states.  This initial state has a total angular
momentum, $F$, that ranges from 1.75 to 2.01.  The final conduction
state is the singly degenerate, lowest energy electron state, with an
$s$-like envelope function and a total angular momentum that ranges
from 1.11 to 1.27.  In the approximate k.p language, this transition is
closest to the $S_{3/2}$ to $S_{1/2}$ transition.

(b) ``peak b'' corresponds to a transition with very weak intensity.
For the largest dot with 42.2\AA~diameter, this state has merged with
peak (a) and cannot be resolved.  The initial valence state associated
with this peak is a doubly degenerate hole state with $p$ character
derived from both heavy and light hole bulk states.  It is the second
highest energy hole state.  This initial state has a total angular
momentum that ranges from 1.92 to 2.51.  The final conduction state is
the same as in peak (a).  In the approximate k.p language this
transition is closest to the $P_{5/2}$ to $S_{1/2}$ transition.

(c) ``peak c'' has a strong intensity.  The initial valence state
associated with this peak is a doubly degenerate hole state with a mix
of $s$, $p$ and $d$ character.  Its most significant contribution is
from the split-off state but it contains some heavy and light hole
character.  This initial state has a total angular momentum that
ranges from 1.99 to 2.48.  The final conduction state is the same as
in peak (a).  In the approximate k.p language this transition is
closest to the $P_{5/2}$ to $S_{1/2}$ transition.

(d) ``peak d'' is a transition with very weak intensity.  For the
largest dot with 42.2\AA~diameter this transition merges with ``peak
c'' and is no longer distinguishable.  The initial valence state
associated with this peak is a doubly degenerate hole states with a
mix of mostly $s$ and $d$ character.  It is derived from the bulk
heavy and light hole states.  This initial state has a total angular
momentum that ranges from 2.48 to 3.25.  The final conduction state is
the same as in peak (a).  In the approximate k.p language this
transition is closest to the $S_{5/2}$ to $S_{1/2}$ transition.

(e) ``peak e'' has a similar origin to ``peak d'' and also has a very
weak intensity.  It also merges with ``peak c'' in the largest dot
with 42.2\AA~diameter.  The initial valence state associated with this
peak is a singly degenerate hole state with a mix of $s$, $d$ and some
$g$ character.  It is derived from both the bulk spin-orbit, heavy and
light hole states.  This initial state has a total angular momentum
that ranges from 3.03 to 3.75.  The final conduction state is the same
as in peak (a).  In the approximate k.p language this transitions is
closest to the $S_{7/2}$ to $S_{1/2}$ transition.

(f) ``peak f'' corresponds to a transition with very weak intensity
that is only observed in the two smaller dots.  It has the same
initial state as ``peak a'', but the final state is the next highest
conduction state.  This conduction state is a triply degenerate state,
with a $p$-like envelope function and a total angular momentum that
ranges from 2.15 to 2.19.  In the approximate k.p language, this
transitions is closest to the $S_{3/2}$ to $P_{3/2}$ transition.

(g) ``peak g'' corresponds to a transition from the same initial state
as ``peak b'', to the $p$-like conduction state.  This is a very
intense peak that is observed in all the sizes of dot.  In the
approximate k.p language, this transitions is closest to the $P_{5/2}$
to $P_{3/2}$ transition.

(h) ``peak h'' is also a very intense peak that is observed in all the
sizes of dot.  The initial valence states are a combination of the
same initial state as ``peak c'' and also a doubly degenerate hole
state, very close in energy to this state with mostly $s$ and $f$
character, derived from heavy and light hole states, with a total
angular momentum ranging from 2.66 to 2.72.  The final conduction
state is the same as in peak (f).  In the approximate k.p language,
this transition is closest to the $P_{5/2}$ to $P_{3/2}$ transition.
It is possible that the proximity of peaks (g) and (h) combined with
their strong intensity would not allow them to be distinguished in a
photoluminescence experiment.  This is discussed further in
section~\ref{ensemble_results}.

(i) ``peak i'' is has a weaker intensity, but is observed in all the
sizes of dot.  The initial valence state is the same as ``peak d''.
The final conduction state is the same as in peak (f).  In the
approximate k.p language, this transition is closest to the $S_{5/2}$
to $P_{3/2}$ transition.  This transition is weakly allowed.

(j) ``peak j'' corresponds to a transition with strong intensity.  The
initial valence state associated with this peak is a doubly degenerate
hole state containing mostly a mix of $p$, $d$ and $f$ character.  Its
contains significant contributions from both the bulk split-off state
and the heavy and light hole states.  This initial state has a total
angular momentum that ranges from 3.03 to 3.41.  The final conduction
state is the same as in peak (f).  n the approximate k.p language,
this transition is closest to the $S_{7/2}$ to $P_{3/2}$ transition.

(k) ``peak k'' corresponds to a transition with weak intensity.  The
initial valence state associated with this peak is a doubly degenerate
hole state containing mostly $p$ character.  It is mostly derived from
the bulk heavy and light hole states.  This initial state has a total
angular momentum that ranges from 3.19 to 3.24.  The final conduction
state is the same as in peak (f).  In the approximate k.p language,
this transition is closest to the $P_{7/2}$ to $P_{3/2}$ transition.

Having discussed the identities of the peaks in each dot size, we next
wish to see how to connect peaks with the same identities in different
sizes of dot.  This is not always possible, as different sizes of dot
might have some peaks that are fundamentally new, or two peaks that
have merged together.  We have labelled in Fig.~\ref{spectra_fig} and
in Table~\ref{peaks_table_12} the peaks that originate from similar
excitons in the different sizes of dot by the same letters (a) to (k).

\subsection{Comparison of calculated single-dot spectra with the
experimental absorption spectra} 

The positions of the above peaks are plotted as a function of diameter
for each dot in Fig.~\ref{absolute} and with respect to the band gap
of each dot in Fig.~\ref{peaks}.  As some of the peaks merge with each
other in the larger dots, not all the peaks are marked for all the
sizes of dot.  The predictions of Figs.~\ref{absolute} and \ref{peaks}
pertain to hypothetical samples each containing dots of a single size
and spherical shape.  The actual synthesized
samples\cite{alivisatos98} contain a significant, but unknown,
distribution of dot sizes and shapes.  Nevertheless, the experimental
results from Ref.\onlinecite{alivisatos98} are included in
Figs.~\ref{absolute} and \ref{peaks}.  It should be noted that where
two peaks have effectively merged together, or where the weight of a
particular peak is too small to be detected (e.g. peak ``b''), these
points are not plotted in Figs.~\ref{absolute} and \ref{peaks}.  A
comparison of the pseudopotential results in Fig.\ref{peaks} with the
results from Ref.\onlinecite{alivisatos98} is given in
Table~\ref{comparison}, which shows that
\begin{enumerate}
\item Peak (a), by definition, corresponds to the experimentally
measured band gap, E1.
\item We ascribe two weak sets of peaks(c and d) as originating
from the weakly observed E2 experimental peak.
\item For the E3 peak we calculate a single peak (e).  
\item For the weakly observed E4 peak we find no calculated
counterpart.  
\item The strongest two calculated peaks (g and h) fall on
either side of the E5 peak and it is possible that the strength of
these excitations could prevent them from being isolated in the
experiment.  They are also merged by a finite size distribution (see section~\ref{ensemble_results}).
\item The final two calculated peaks (j and k) correspond to the
experimental E6 and E7 data.  
\end{enumerate}
For each of the peaks, the calculated scaling of the exciton energy
{\em spacings} with dot size (or band gap) shows reasonable agreement
with the experimental results.  However, Fig.~\ref{absolute} shows
that the calculated values of the {\em absolute} exciton peaks appears
to exhibit a different size dependence to that observed in
Ref.\onlinecite{alivisatos98}.  This lack of agreement can be
attributed in part to the finite size and shape distribution present
in the experimental samples.  This is discussed in the following
section.

\subsection{Ensemble absorption spectra}\label{ensemble_results}

The ideal comparison between theory and experiment is between the
calculated (Fig.~\ref{spectra_fig}) and measured {\em single-dot}
spectra.  However, no such single-dot measurements currently exist for
InAs quantum dots.  Our predicted single-dot spectra for different
sizes (Fig.~\ref{spectra_fig}) suggest that the interpretation of an
ensemble spectra could differ qualitatively from the interpretation of
a single dot spectra.  This calls for measurement of the single dot
spectra.  For example, peak (b) in the dot with a diameter of 30.3
\AA~coincides with peak (e) in the dot with a diameter of 36.9 \AA~and
with peak (g) in the dot with a diameter of 42.2\AA.  Thus, if the
experimentally accessible samples represent a broad size distribution
it is impossible to ascribe consistently experimental peaks to unique
calculated single-dot peaks.

We try to quantify the effect of a finite size distribution in the
experimental samples by using our single dot spectra in combination
with Eq.(\ref{ensemble_eq}) to calculate {\em ensemble} absorption
spectra.  These are not directly comparable with the size selected PLE
results from Ref.\onlinecite{banin96}, but are designed to provide a
general indication of the effects of size distribution on the
absorption spectra of an ensemble of dots.  We neglect shape
distribution effects, since they are not quantified experimentally.
Transmission electron microscopy studies of III-V semiconductor
quantum dots\cite{micic97} show that there are two factors producing
an ensemble of different dot volumes.  Firstly, in any sample there is
a finite range of dot diameters.  Secondly, the dots are ellipsoidal
in shape, with a range of ratio of major to minor axes.  We have
therefore chosen to model the distribution of sizes, $P(R)$, in
Eq.(\ref{ensemble_eq}) by a simple Gaussian, whose width, $\sigma_R$,
builds in the size distribution,
\begin{equation}\label{distribution}
P(R) =
\frac{1}{\sqrt{2\pi\sigma_R}}e^{-(R-R_0)^2/2\sigma_R^2} \;\;\; .
\end{equation}
In Fig.~\ref{ensemble_spectra}(a) we plot ensemble absorption spectra
calculated from Eq.(\ref{ensemble_eq}) for quantum dots with a mean
diameter of 23.9 \AA~and standard deviations, $\sigma_R$, of 0, 5 and
10\% of the mean size.  The function $I(E,R)$ in
Eq.(\ref{ensemble_eq}) was obtained by fitting the size dependence of
each of the peaks, $i$, in Fig.~\ref{absolute} to $E_i(R)=E_i^0+aR^N$
and then summing the contributions from all the peaks so that
\begin{equation}
I(E,R) = \sum_{peaks, i} \alpha_i e^{-(E-E_i(R))^2/2\sigma_0^2}
\;\;\; ,
\end{equation}
where $\alpha_i$ is the relative intensity of peak $i$, and $\sigma_0$
is the intrinsic line width of the peaks, set to 5 meV.
Figures~\ref{ensemble_spectra}(b) to (d) show similar ensemble
absorption spectra for dots with size distributions that have mean
diameters of 30.3, 36.9 and 42.2 \AA~and standard deviations of 5\%.
Figures~\ref{ensemble_spectra}(a) to (d) show that as the width of the
size distribution increases, the peaks become ``smeared out''.  With a
standard deviation of 5\% it is still possible to resolve peaks, (a),
(c), (g) and (h).  However, peak (j) is just a small shoulder on peak
(h).  With a standard deviation of 10\% peaks (g) and (h) have
effectively merged together.  The merging effect becomes more severe
as the mean size of the dots increases as the initial spacing of the
peaks is smaller for larger dots.

Figure~\ref{ensemble_peaks} shows how the centers of the broadened
peaks in each of the 5\% ensemble spectra from
Fig.~\ref{ensemble_spectra} vary with size.
Figure~\ref{ensemble_peaks} is the size ensemble equivalent of the
single dot results plotted in Fig.~\ref{peaks}.  As in
Fig.~\ref{peaks}, the experimental results of Banin {\em et. al.} are
plotted for comparison.  Comparison of Figs.~\ref{peaks} and
\ref{ensemble_peaks}, reveals that
\begin{enumerate}
\item The fundamental transition (a) is still clearly resolvable in the
ensemble spectra and corresponds to experimental peak E$_1$.
\item The first excited state (c) is still clearly resolvable in all
sizes of dot.  This peak probably still corresponds to peak E$_2$.
\item Peak (d) is only resolvable in the two smallest dots, where it
is close to the experimental E$_3$ peak.
\item For the larger two dots, peaks (g) and (h) merge to form one
large peak.  The position of this merged peak is close to the
experimental E$_5$ peak.  For the two smaller dots, the peak splits
into two peaks (g) and (h) with different size scaling behaviour.
\item Peak (j) is only resolvable from peaks (g) and (h) in the two
smaller dots, where it could correspond to either the experimental
E$_6$ or E$_7$ peaks.
\item The weaker peaks (b), (e) and (k) are not individually
resolvable for any size of dot in the ensemble spectra.
\end{enumerate}

\section{Conclusions}\label{conclusions}

We have performed pseudopotential calculations for the
electronic structure of both the ground and excited states of free
standing InAs quantum dots for a range of experimentally realistic
sizes.  Using calculated electron-hole Coulomb energies and dipole
matrix transition probabilities we have constructed single-dot
absorption spectra for 4 different sizes of quantum dot.  These
spectra exhibit a series of clearly resolvable exciton peaks.  The
size dependence of the spacing between the exciton peaks in these
single dot spectra shows partial agreement with those found in recent
experiments.  By fitting the size dependence of each exciton peak and
postulating a Gaussian distribution of dot sizes, we have calculated
ensemble absorption spectra.  The size scaling of the peaks in these
ensemble spectra shows a better agreement with the experimental data.

We also have analyzed the single particle parentage of each excitonic
peak.  We find that (i) as a result of the small band gap of InAs,
there is significant valence-conduction band mixing in the quantum dot
states, (ii) the removal of spherical symmetry of these dots produces
odd-even mixing in these states.

Our predicted single-dot excitonic spectra (Fig.~\ref{spectra_fig})
await experimental testing.  Our predicted ensemble spectra
(Figs.~\ref{ensemble_spectra}) are not in as good agreement with
experiment as our results for InP\cite{fu98} and CdSe\cite{wang_cdse}
dots.  Sample quality, including shape distributions could be a factor
in this relative lack of agreement.

\noindent{\bf Acknowledgements} We thank L.W. Wang and
A. Franceschetti for useful discussions and their comments on the
manuscript.  This work was supported DOE -- Basic Energy Sciences,
Division of Materials Science under contract No. DE-AC36-98-GO10337.

%\bibliographystyle{i:/latex/phaip}
%\bibliography{i:/latex/GaInP}

\begin{thebibliography}{10}

\bibitem{mrs_feb98}
A.~Zunger,
\newblock MRS Bulletin {\bf 23}, 35 (1998).

\bibitem{samuelson94}
N.~Carlsson, W.~Seifert, A.~Petersson, P.~Castrillo,
M.E.~Pistol and L.~Samuelson, 
\newblock Appl. Phys. Lett {\bf 66}, 3093 (1994).

\bibitem{petroff96}
K.~Schmidt, G.~Medeiros-Ribeiro, M.~Oestreich, and P.~Petroff,
\newblock Phys. Rev B. {\bf 54}, 11346 (1996).

\bibitem{grundman95}
M.~Grundmann, O.~Stier, and D.~Bimberg,
\newblock Phys. Rev. B {\bf 52}, 11969 (1995).

\bibitem{bawendi96}
S.~Empedocles, D.~Norris, and M.~Bawendi,
\newblock Phys. Rev. Lett. {\bf 77}, 3873 (1996).

\bibitem{banin96}
A.~Guzelian, U.~Banin, A.~Kadavanich, X.~Peng, and A.~Alivisatos,
\newblock Appl. Phys. Lett. {\bf 69}, 1432 (1996).

\bibitem{micic96}
O.~Micic, J.~Sprague, Z.~Lu, and A.~Nozik,
\newblock Appl. Phys. Lett. {\bf 68}, 3150 (1996).

\bibitem{bawendi96:2}
D.~Norris and M.~Bawendi,
\newblock Phys. Rev. B {\bf 53}, 16338 (1996).

\bibitem{colvin91}
A.~Colvin, V.L.and~Alivisatos and J.~Tobin,
\newblock Phys. Rev. Lett. {\bf 66}, 2786 (1991).

\bibitem{micic98}
D.~Bertram, O.~Micic, and A.~Nozik,
\newblock Phys. Rev. B {\bf 57}, R4265 (1998).

\bibitem{alivisatos98}
U.~Banin, J.C.~Lee, A.A~Guzelian, V.~Kadavanich, A.P.~Alivisatos,
W.~Jaskolski, G.Q.~Bryant, Al.L.~Efros and M.~Rosen
\newblock J. Chem. Phys. {\bf 109}, 2306.

\bibitem{wang_cdse}
L.-W. Wang and A.~Zunger,
\newblock J. Phys. Chem. {\bf 102}, 6449 (1998).

\bibitem{fu98}
H.~Fu and A.~Zunger,
\newblock Phys. Rev. B {\bf 57}, R15064 (1998).

\bibitem{wang96}
L.-W. Wang and A.~Zunger,
\newblock Phys. Rev B. {\bf 53}, 9579 (1996).

\bibitem{mizel98}
A.~Mizel and M.~Cohen,
\newblock Solid State Commun. {\bf 104}, 401 (1998).

\bibitem{zhang93}
S.~Zhang and A.~Zunger,
\newblock Phys. Rev. B {\bf 48}, 11204 (1993).

\bibitem{williamson98:2}
A.~Williamson and A.~Zunger,
\newblock Phys. Rev. B {\bf 58}, 6724 (1998).

\bibitem{jkim98}
J.~Kim, L.-W. Wang, and A.~Zunger,
\newblock Phys. Rev. B {\bf 57}, R9408 (1998).

\bibitem{williamson98:3}
A.~Williamson and A.~Zunger,
\newblock Phys. Rev. B , Code: BU6515 (1998).

\bibitem{wang98}
L.-W. Wang, J.~Kim, and A.~Zunger,
\newblock Phys. Rev B. {\bf 59}, 5678 (1999).

\bibitem{note1}
 {\bf In practice, we use two artificial barrier materials, one for valence
  band calculations, and one for the conduction band. The use of two separate
  potentials enables us to achieve larger valence and conduction band offsets
  (VBO and CBO), between the dot and the barrier, than could be realized with a
  single artificial barrier material. We have a VBO of 2.1 eV and a CBO of 3.2
  eV. The fitted values for the artificial barrier are given in
  Table~\ref{values_table} Although, in principle, the choice of different
  barrier potentials means the valence and conduction states are no longer
  orthogonal, in practice the difference in the barrier potentials is small and
  there is also minimal spillage of the wavefunctions into this barrier
  material. The resulting overlap between valence and conduction states is
  typically only 3x10$^{-5}$.}

\bibitem{wang95}
L.-W. Wang and A.~Zunger,
\newblock Phys. Rev B. {\bf 51}, 17398 (1995).

\bibitem{wang94}
L.-W. Wang and A.~Zunger,
\newblock J. Chem. Phys. {\bf 100}, 2394 (1994).

\bibitem{kamat}
L.-W. Wang and A.~Zunger,
\newblock {\em Semiconductor Nanoclusters},
\newblock Elsevier Science, Amsterdam, 1996.

\bibitem{franceschetti99}
A.~Franceschetti, H.~Fu, L.-W. Wang, and A.~Zunger,
\newblock Phys. Rev. B. {\bf 60}, 1819 (1999).

\bibitem{resta77}
R.~Resta,
\newblock Phys. Rev. B {\bf 16}, 2717 (1977).

\bibitem{haken}
H.~Haken,
\newblock Nuovo Cimento {\bf 10}, 1230 (1956).

\bibitem{tsu}
R.~Tsu, L.~Ioriatti, J.~Harvey, H.~Shen, and R.~Lux,
\newblock Mater. Res. Soc. Symp. Proc. {\bf 283}, 437 (1993).

\bibitem{penn}
D.~Penn,
\newblock Phys. Rev {\bf 128}, 2093 (1962).

\bibitem{brus86}
L.~Brus,
\newblock J. Phys. Chem. {\bf 90}, 2555 (1986).

\bibitem{franceschetti97}
A.~Franceschetti and A.~Zunger,
\newblock Phys. Rev. Lett. {\bf 78}, 915 (1997).

\bibitem{banin98}
U.~Banin, C.J.~Lee, A.A~Guzelian, A.V.~Kadavanich, A.P.~Alivisatos,
W.~Jaskolski, G.W.~Bryant, Al.L.~Efros and M.~Rosen
\newblock J. Chem. Phys. {\bf 109}, 2306 (1998).

\bibitem{banin99}
U.~Banin, D.~Katz, and O.~Millo,
\newblock Nature , accepted for publication (1999).

\bibitem{van_buuren}
T.~van Buuren, L.~Dinh, L.~Chase, W.~Siekhaus, and L.~Terminello,
\newblock Phys. Rev. Lett. {\bf 80}, 3803 (1998).

\bibitem{web_site}
 {\bf More details can be found in the ``downloadable data'' section of {\tt
  http://www.sst.nrel.gov}}.

\bibitem{micic97}
O.~Micic, H.M.~Cheong, H.~Fu, A.~Zunger, J.R.~Sprague, A.~Mascarenhas
and A.J.~Nozik 
\newblock J. Phys. Chem. B {\bf 101}, 4904 (1997).

\bibitem{bornstein}
Landolt and Borstein,
\newblock {\em Numerical Data and Functional Relationships in Science and
  Technology, Volume 22, Subvolume a},
\newblock Springer-Verlag, Berlin, 1997.

\bibitem{fu97}
H.~Fu and A.~Zunger,
\newblock Phys. Rev. B {\bf 55}, 1642 (1997).

\bibitem{reboredo99}
F.Reboredo and A.~Zunger,
\newblock Phys. Rev. B , BE7134.

\end{thebibliography}

\newpage
\begin{table}[hbt]
\caption{Experimental and fitted bulk InAs pseudopotential
properties.  $\Delta_{SO}$ is the spin-orbit splitting and
$m^*_{\Gamma_{1c}}$, $m^*_{\Gamma_{15v},hh}$ anFig.~\ref{gaps}(a)d
$m^*_{\Gamma_{15v},lh}$ are the effective masses of electrons, heavy
and light holes.}
\label{fit_results}
\begin{tabular}{ccc}
Property & Expt. Value\cite{bornstein} & Pseudopotential \\
\hline
$\Gamma_{1c}-\Gamma_{15v}$ (eV) & 0.42 & 0.41 \\
$X_{1c}-\Gamma_{15v}$ (eV) & 2.33 & 2.27 \\
$L_{1c}-\Gamma_{15v}$ (eV) & 1.71 & 1.61 \\
$\Delta_0$ (eV) & 0.38 & 0.36 \\
$m^*_{\Gamma_{1c}}$ & 0.029 & 0.028 \\
$m^*_{\Gamma_{15v},hh}[100]$ & 0.43 & 0.41 \\
$m^*_{\Gamma_{15v},lh}[100]$ & 0.038& 0.039 \\
\end{tabular}
\end{table}

\newpage
\begin{table}[hbt]
\caption{Screened atomic pseudopotential parameters for InAs and the
barrier potentials.  The $a$ parameters refer to Eq.(\ref{epm}).}
\label{values_table}
\begin{tabular}{cccccc}
& $a_0$ & $a_1$ & $a_2$ & $a_3$ \\ 
\tableline 
In & 118.781 & 1.783 & 3.382  & 0.393 \\
Barrier Cation (cond) & 107.755 & 1.915 & 3.460 & 0.414 \\
Barrier Cation (val) & 333.070 & 0.120 & 3.126 & 0.521 \\
\hline
As & 65.943 & 2.664 & 1.684 & 0.637 \\
Barrier Anion (cond) & 19.892 & 2.097 & 1.182 & 0.243 \\
Barrier Anion (val) & 49.614 & 2.737 & 1.523 & 0.574 \\
\end{tabular}
\end{table}

\newpage
\begin{table}[hbt]
\caption{InAs quantum dot sizes and compositions}
\label{dots}
\begin{tabular}{ccccc}
Dot Number & 1 & 2 & 3 & 4 \\
\hline
No. In atoms & 140 & 276 & 456 & 736 \\
No. As atoms & 141 & 249 & 459 & 683 \\
Diameter (\AA) & 23.9 & 30.3 & 36.6 & 42.2 \\
\end{tabular}
\end{table}

\newpage
\begin{table}[hbt]
\caption{A comparison of the size scaling exponent, $n$, in $R^{-n}$
of the unscreened electron-hole Coulomb energy ,$\epsilon J$, the
screened electron-hole Coulomb energy, $J$, the single particle band
gap, $E_{gap}^{sp}$ and the excitonic band gap $E_{gap}^{exciton}$
[Eq.(\ref{exciton_energy})] for InAs, InP, CdSe and Si quantum dots.}
\label{scalings}
\begin{tabular}{cccccc}
Property & InAs & InP\cite{fu97} & CdSe\cite{wang_cdse} & Si\cite{reboredo99}  & EMA \\
\hline
$\epsilon J$ & 0.70 & & 0.86 & 0.82 & 1 \\
$J$ & 1.79 & 1.35 & 1.18 & 1.50 & 1 \\
$E_{gap}^{sp}$ & 1.01 & 1.16 & 1.64 & 1.20 & 2 \\
$E_{gap}^{exciton}$ & 0.90 & 1.09 & 1.93 & 1.18 \\
\end{tabular}
\end{table}

\newpage
\begin{table}[hbt]
\caption{Contributions (as fractions of unity) from the split-off
(SO), heavy hole (hh), light hole (lh) and conduction band (CBM) bulk
states to the single particle states of the InAs dot with
42.2\AA~diameter.  The dot states are numbered (first column) from the
band edge.  For electron states 1 is the lowest in energy, 8 the
highest.  For hole states, 1 is the highest in energy, 24 the
lowest. The contribution from each state is determined using
Eq.(\ref{decomp}).  For each state the bulk SO, hh+lh and CBM contributions
are decomposed into their angular momentum components using
Eq.(\ref{weight}).  Only the states contributing to the major
excitonic peaks in Fig.~\ref{spectra_fig} are shown.  All entries less
than 0.01 are set to 0.0.  The main contributions are marked in bold.
Totals include envelope functions with angular momenta from 0 to 6.}
\label{analysis_table_12}
\begin{tabular}{cc|cccc|cccc|cccc}
&& \multicolumn{4}{c|}{SO Contribution} &
\multicolumn{4}{c|}{hh + lh Contribution} &
\multicolumn{4}{c}{CBM Contribution} \\
\cline{3-6}\cline{7-10}\cline{11-14}
Level & Energy (eV) & $s$ & $p$ & $d$ & Total & $s$ & $p$ & $d$ & Total &
$s$ & $p$ & $d$ & Total \\ 
\hline
\multicolumn{2}{c|}{{\bf Conduction States}}&&&&&&&&&&&\\
1 & $\epsilon_c$ & 0. & 0.07 & 0. & 0.07 & 0. & {\bf 0.20} &
0. & {\bf 0.20} & {\bf 0.69} & 0. & 0. & {\bf 0.69} \\
2 & $\epsilon_c+0.360$ & 0.06 & 0. & 0.01 & 0.07 & 0.01 & 0.
& {\bf 0.19} & {\bf 0.20} & 0. & {\bf 0.64} & 0.01 & {\bf 0.65} \\
3,4 & $\epsilon_c+0.361$ & 0. & 0. & 0.07 & 0.07 & 0.09 &
0. & {\bf 0.10} & {\bf 0.19} & 0. & {\bf 0.64} & 0.01 & {\bf 0.65} \\
\multicolumn{2}{c|}{{\bf Valence States}}&&&&&&&&&&&\\
1,2 & $\epsilon_v$ & 0. & 0.02 & 0.02 & 0.04 & {\bf 0.50} & {\bf 0.29} &
0.09 & {\bf 0.89} & 0. & 0.02 & 0. & 0.02 \\
3,4 & $\epsilon_v-0.014$ & 0. & 0.01 & 0.01 & 0.03 & {\bf 0.23} &
{\bf 0.55} & 0.07 & {\bf 0.88} & 0. & 0.01 & 0. & 0.01 \\
7,8 & $\epsilon_v-0.098$ & 0. & 0.02 & 0.02 & 0.08 & 0.08 &
{\bf 0.38} & {\bf 0.18} & {\bf 0.78} & 0. & 0. & 0.02 & 0.02 \\
11,12 &  $\epsilon_v-0.123$ & {\bf 0.28} & 0. & 0. & {\bf 0.31} & 0. &
{\bf 0.26} & {\bf 0.19} & {\bf 0.56} & 0. & 0. & 0.01 & 0.01 \\
13 & $\epsilon_v-0.140$ & 0. & {\bf 0.21} & 0.06 & {\bf 0.28} & 0. &
0.12 & {\bf 0.21} & {\bf 0.60} & 0. & 0. & 0. & 0. \\
15,16 & $\epsilon_v-0.175$ & 0. & 0.06 & 0.01 & 0.14 & 0.02 &
{\bf 0.26} & 0.05 & {\bf 0.73} & 0. & 0. & 0. & 0. \\
18 & $\epsilon_v-0.189$ & 0. & 0.04 & {\bf 0.13} & {\bf 0.27} & 0.07 &
{\bf 0.16} & 0.04 & {\bf 0.69} & 0. & 0. & 0. & 0. \\
23,24 & $\epsilon_v-0.223$ & 0. & 0.05 & 0.01 & 0.14 & 0.01 &
{\bf 0.33} & {\bf 0.12} & {\bf 0.69} & 0. & 0. & 0.02 & 0.02
\end{tabular}
\end{table}

\newpage
\begin{table}[hbt]
\caption{Analysis of the significant absorption peaks for a dot with
42.2\AA~diameter.  The peak letters refer to the labelling of the
peaks in Figs.~\ref{peaks} and \ref{spectra_fig}.  Peak energies are
calculated according to Eq.(\ref{exciton_energy}).  The valence and
conduction indices refer to the number of the state from the band edge
(i.e. VBM and CBM have index 1 etc).  The envelope function
information summarizes the results from
Table~\ref{analysis_table_12}.  Note, the notation ``s'', ``p'' etc
refers to the orbital content of the wavefunctions, not the specific
number of nodes. All energies are in eV.}
\label{peaks_table_12}
\begin{tabular}{cccccccccc}
Peak & Peak Energy & \multicolumn{4}{c}{Initial valence state} &
\multicolumn{4}{c}{Final conduction state} \\
\cline{3-6}\cline{7-10}
& & Index & Energy & Envelope & F &
Index & Energy & Envelope & F \\
\tableline 
(a)  & 1.209 & 1,2   & -5.8260 & $s+p$   & 1.77 & 1     & -4.516 & $s$ & 1.11 \\
(c)  & 1.335 & 11,12    & -5.9491 & $p+d$   & 2.48 & 1     & -4.516 & $s$ & 1.11 \\
(g)  & 1.588 & 3,4   & -5.8401 & $p$     & 1.90 & 2,3,4 & -4.156 & $p$ & 1.92	\\
(h)  & 1.675 & 7,8   & -5.9240 & $s+p+f$ & 2.66 & 2,3,4 & -4.156 & $p$ & 1.92	\\
(i)  & 1.717 & 13    & -5.9660 & $p+d+f$ & 2.93 & 2,3,4 & -4.156 & $p$ & 1.92 \\
(j)  & 1.760 & 15,16 & -6.0087 & $p+f$   & 3.07 & 2,3,4 & -4.156 & $p$ & 1.92\\
(k)  & 1.800 & 23,24 & -6.0491 & $p$   & 3.19 & 2,3,4 & -4.156 & $p$ & 1.92\\
\end{tabular}
\end{table}

\newpage
\begin{table}[hbt]
\caption{Comparison of the current multi-band pseudopotential
assignment and the k.p assignment from
Ref.\protect\onlinecite{alivisatos98} of the experimentally observed
optical transitions in Ref.\protect\onlinecite{alivisatos98}.  The
initial states and final states of each transition are denoted by
$nQ_F$, where $n$ is the principal quantum number, $Q$ the lowest
angular momentum component of the wavefunction and $F$ total angular
momentum.  Note, it is only possible to determine the envelope
function angular momentum and total angular momentum for the
multi-band pseudopotential wavefunctions. }
\label{comparison}
\begin{tabular}{cccccccccc}
Peak & Pseudopotential assignment & Closest Expt. Peak & k.p assignment \\
&& from Ref.\onlinecite{alivisatos98} & from
Ref.\onlinecite{alivisatos98} \\
\hline
(a) & $S_{3/2}\rightarrow S_{1/2}$ & E$_1$ & $2S_{3/2}\rightarrow
1S_{1/2}$ \\
(b) & $P_{5/2}\rightarrow S_{1/2}$ & Not observed & No prediction \\
(c) & $P_{5/2}\rightarrow S_{1/2}$ & E$_2$/E$_3$ & No prediction \\
(d) & $S_{5/2}\rightarrow S_{1/2}$ & E$_2$ & No prediction \\
(e) & $S_{7/2}\rightarrow S_{1/2}$ & E$_3$ & $2S_{3/2}\rightarrow
1S_{1/2}$ \\
(g) and (h) & $P_{5/2}\rightarrow P_{3/2}$ & E$_5$ & $\begin{array}{l} 
1P_{3/2}\rightarrow 1P_{3/2} \\ 1P_{3/2}\rightarrow 1P_{1/2}
\end{array}$ \\
(j) & $S_{7/2}\rightarrow P_{3/2}$ & E$_6$ & $2S_{1/2}\rightarrow
1S_{1/2}$ \\
(k) & $P_{7/2}\rightarrow P_{3/2}$ & E$_7$ & $\begin{array}{l} 
1P_{1/2}\rightarrow 1P_{3/2} \\
1P_{1/2}\rightarrow 1P_{1/2} \end{array} $

\end{tabular}
\end{table}

\begin{figure}
\caption{InAs bulk band structure from the $L$ to $\Gamma$ to $X$
calculated with the empirical pseudopotential defined in Eq.(\ref{epm}).}
\label{bulk}
\end{figure}

\begin{figure}
\caption{The electronic and ionic contributions to the dielectric
screening function.  Plotted in reciprocal space.}
\label{dielectric_recip}
\end{figure}

\begin{figure}
\caption{The screened dielectric function.  Plotted in real space for
each of the 4 InAs dots given in Table~\ref{dots}.}
\label{dielectric_real}
\end{figure}

\begin{figure}
\caption{Electron-hole Coulomb energies calculated using both
pseudopotential [Eq.(\ref{Coulomb_energy})] and effective mass
[Eq.(\ref{brus_coulomb})] expressions for all 4 sizes of InAs quantum
dot given in Table~\ref{dots}.}
\label{coulomb_fig}
\end{figure}

\begin{figure}
\caption{Highest energy hole and lowest energy electron wavefunctions
squared in a plane through the center of the dot for a free standing
spherical InAs quantum dot with diameter of 42.2 \AA.}
\label{contour}
\end{figure}

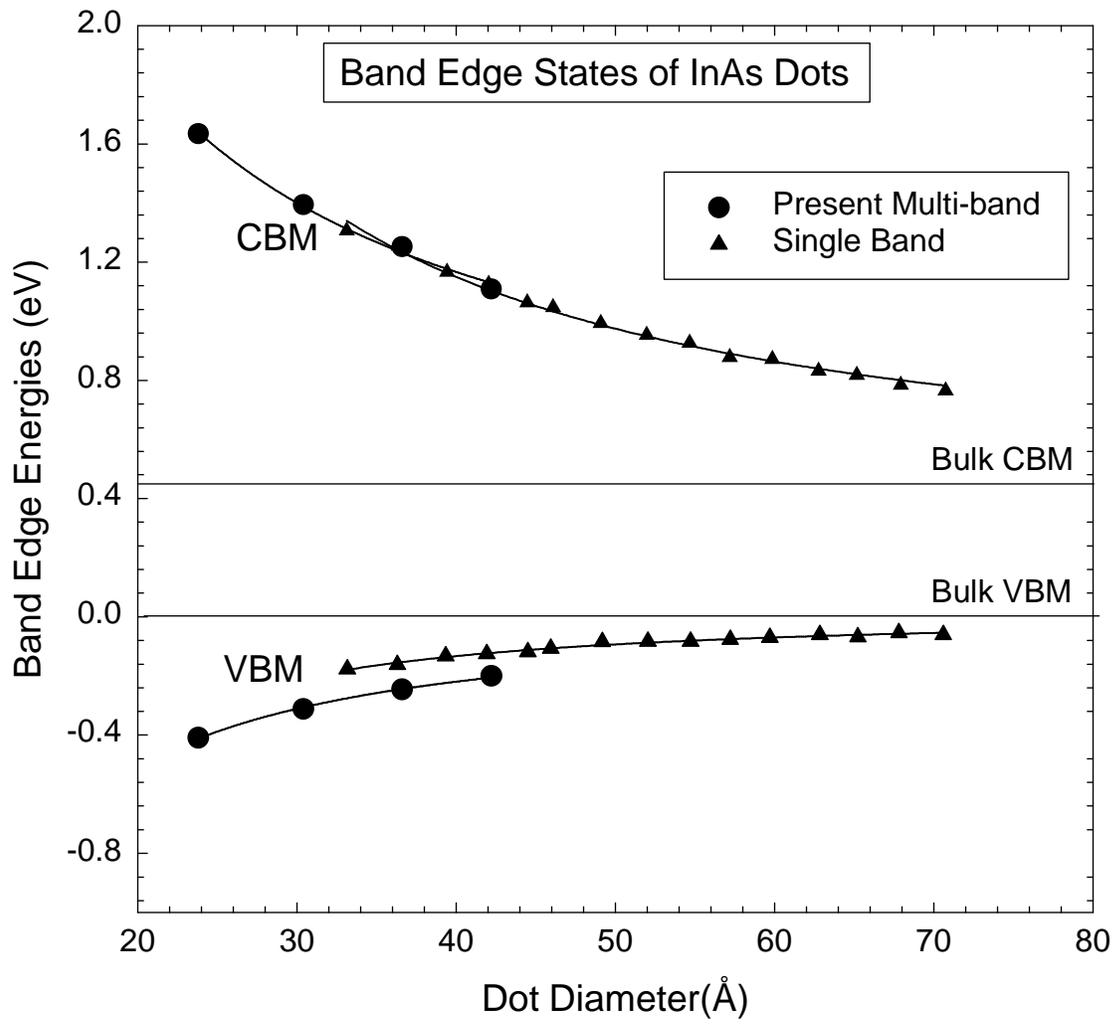
\begin{figure}
\caption{Size dependence of the lowest energy hole state and highest
energy electron states, calculated using the present multi-band
pseudopotential and single band techniques.  The circles and triangles
mark the present multi-band pseudopotential and single band (Wannier
function) results.  The solid lines are the best fit to a scaling of
$E_0+AR^n$ (see Eq.(\ref{scaling})).}
\label{vbm-cbm}
\end{figure}

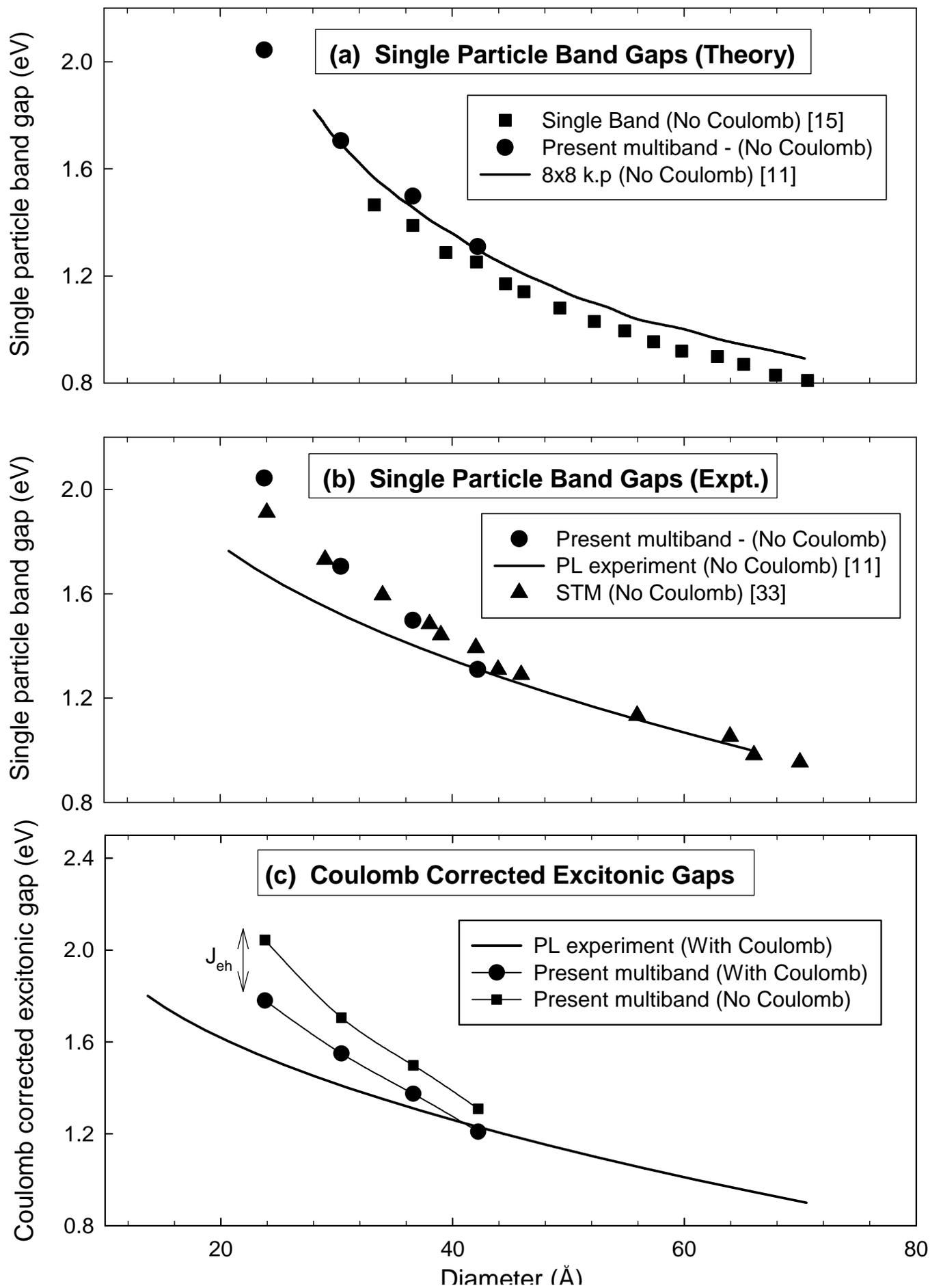
\begin{figure}
\caption{(a) Calculated single particle band gaps calculated using
present multi-band pseudopotential (circles), 8x8 k.p (solid line) and
single band techniques (squares).  (b) Measured single particle band
gaps using Photoluminescence (solid line) and STM experiments
(triangles).  (c) Present multi-band pseudopotential results (circles)
including the electron-hole Coulomb energy calculated from
Eq.(\ref{Coulomb_energy}) vs PL data.}
\label{gaps}
\end{figure}

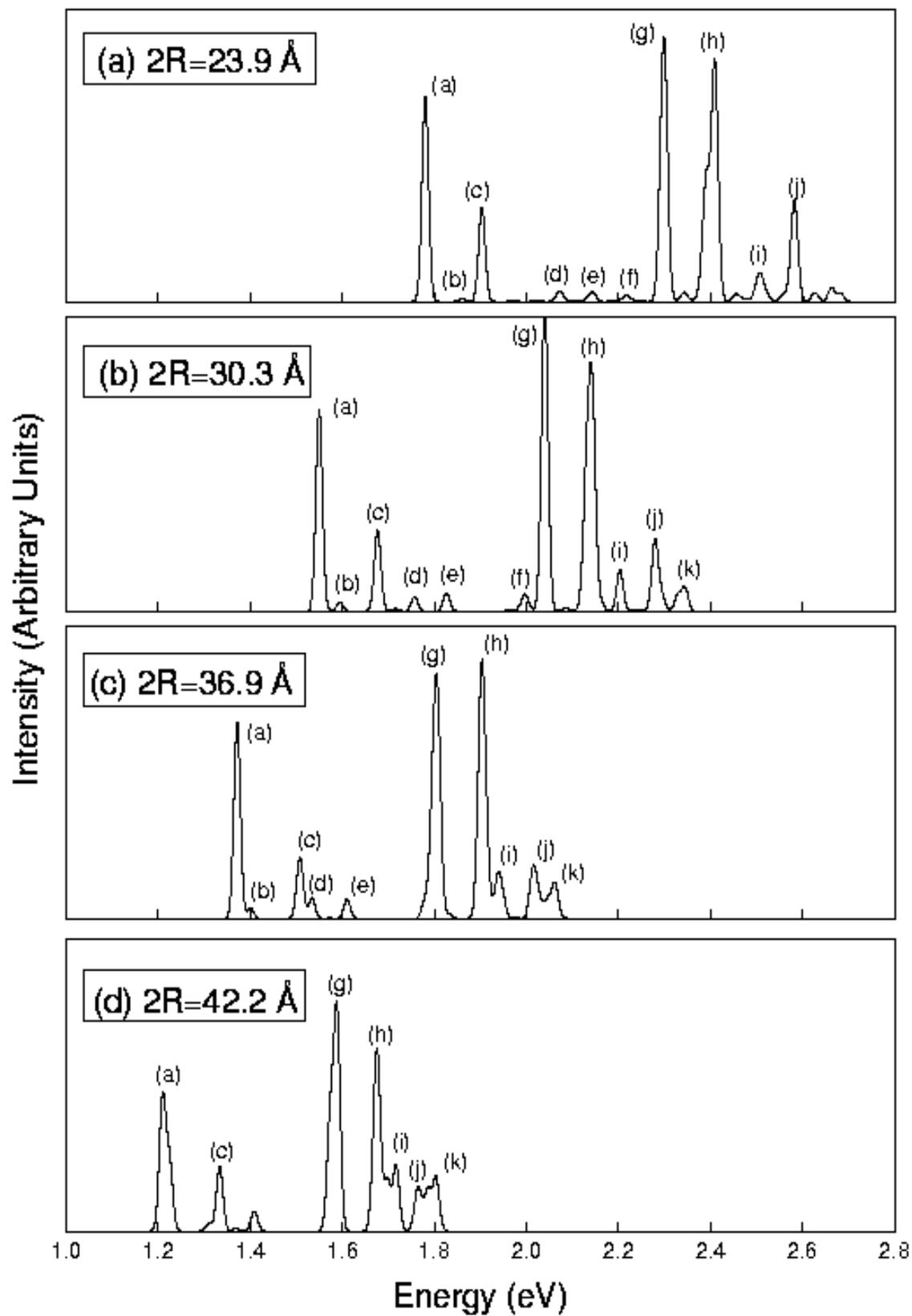
\begin{figure}
\caption{Calculated absorption spectra for single InAs quantum dots
with diameters of (a) 23.9, (b) 30.3, (c) 36.6 and (d) 42.2\AA.  The
labels (a)-(k) refer to the classification of the states in
Table~\ref{peaks_table_12}.}
\label{spectra_fig}
\end{figure}

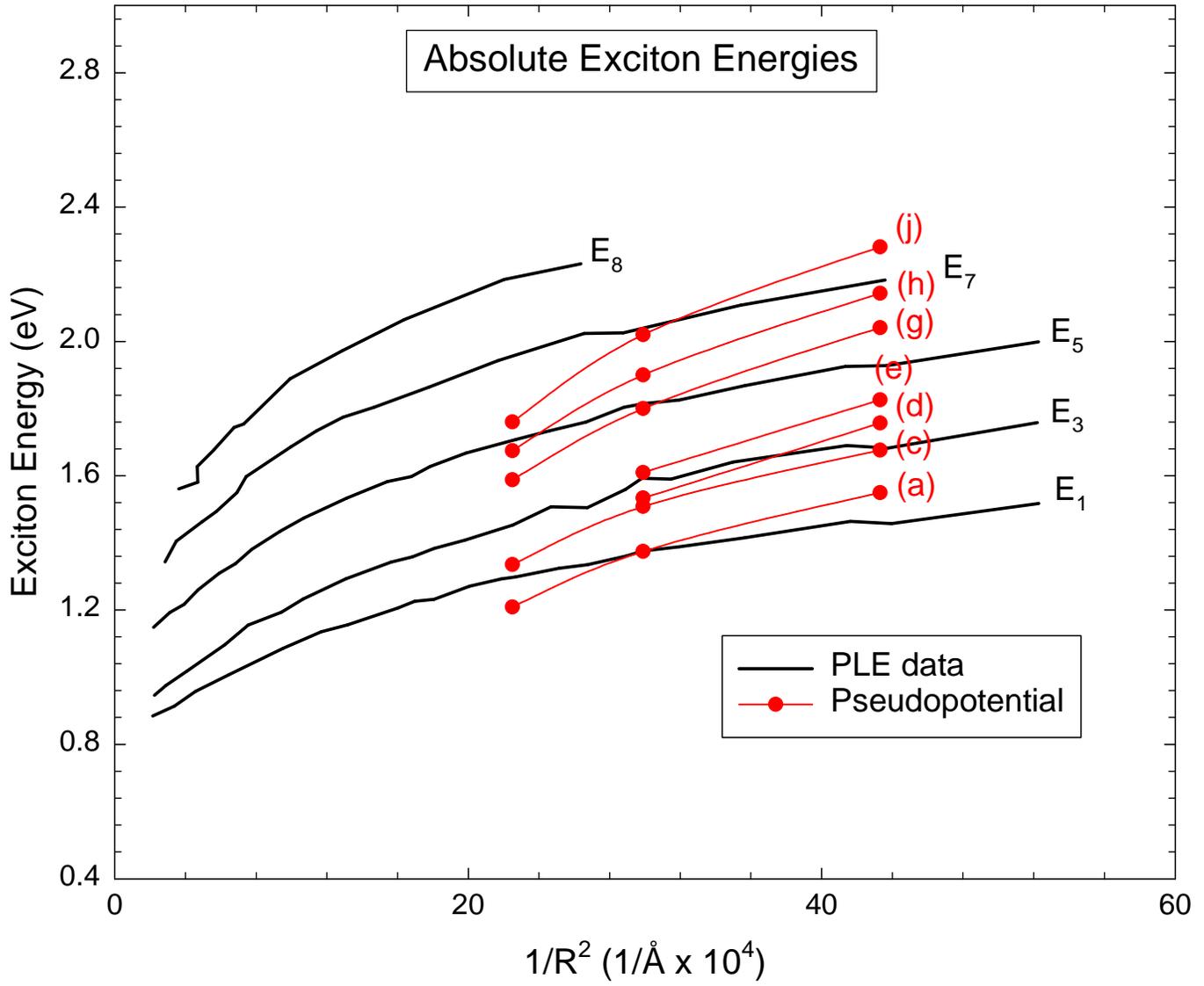
\begin{figure}
\caption{Pseudopotential calculated (circles) absolute positions of
absorption peaks $vs$ $1/R^2$ for InAs quantum dots with diameters
23.9, 30.3, 36.6 and 42.2\AA.  The pseudopotential results are
connected using the analysis from section~\ref{spectra_results}.
Experimental results from Ref.\protect\onlinecite{alivisatos98} are
represented by the black lines.}
\label{absolute}
\end{figure}

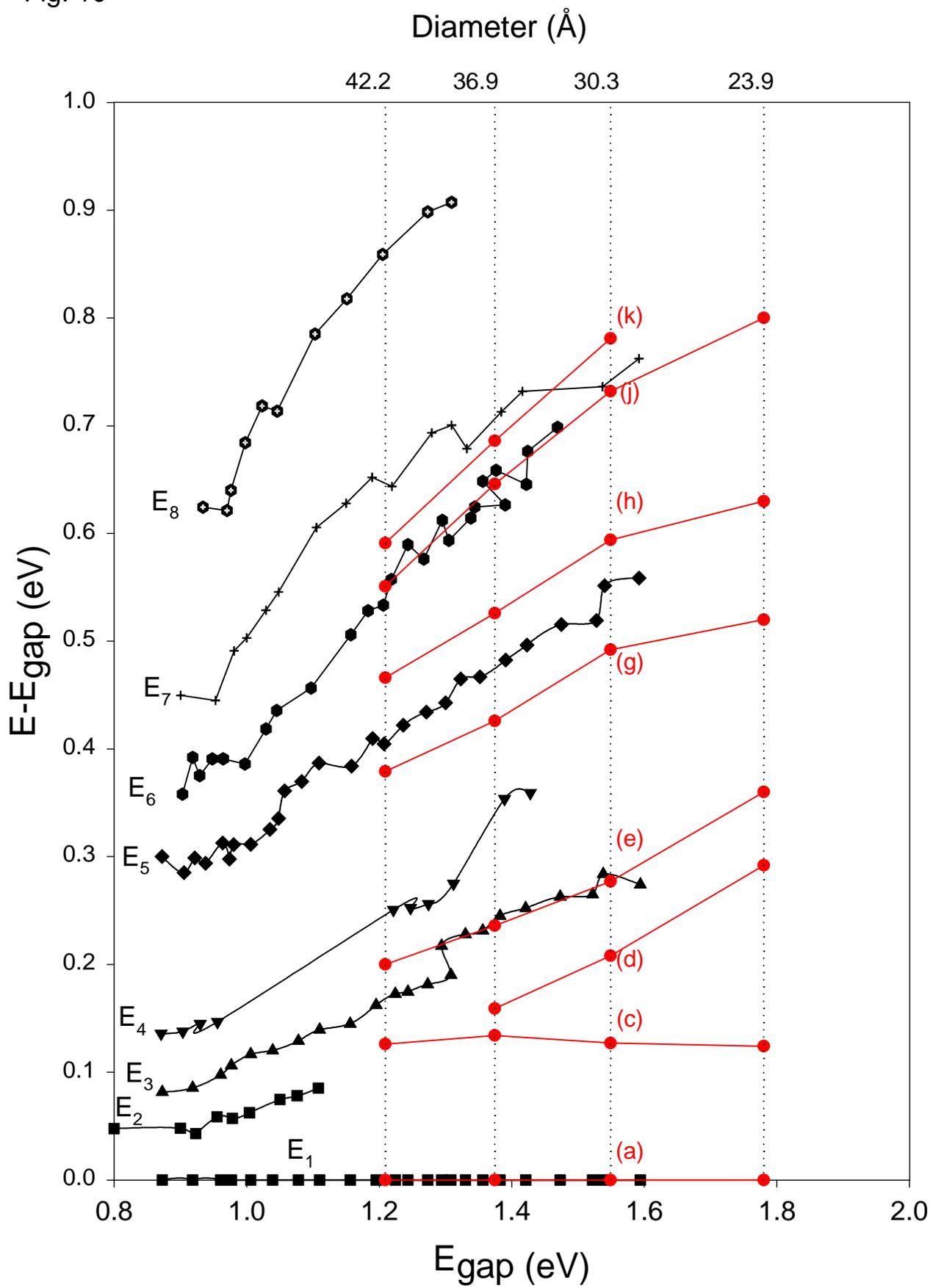
\begin{figure}
\caption{Pseudopotential calculated (circles) positions of absorption
peaks minus the band gap $vs$ band gap for InAs quantum dots with
diameters 23.9, 30.3, 36.6 and 42.2\AA.  The pseudopotential results
are connected using the analysis from section~\ref{spectra_results}.
The labels (a)-(k) refer to the classification of the states in
Table~\ref{peaks_table_12}.  Experimental results from
Ref.\protect\onlinecite{alivisatos98} are represented by the black
symbols and lines.}
\label{peaks}
\end{figure}

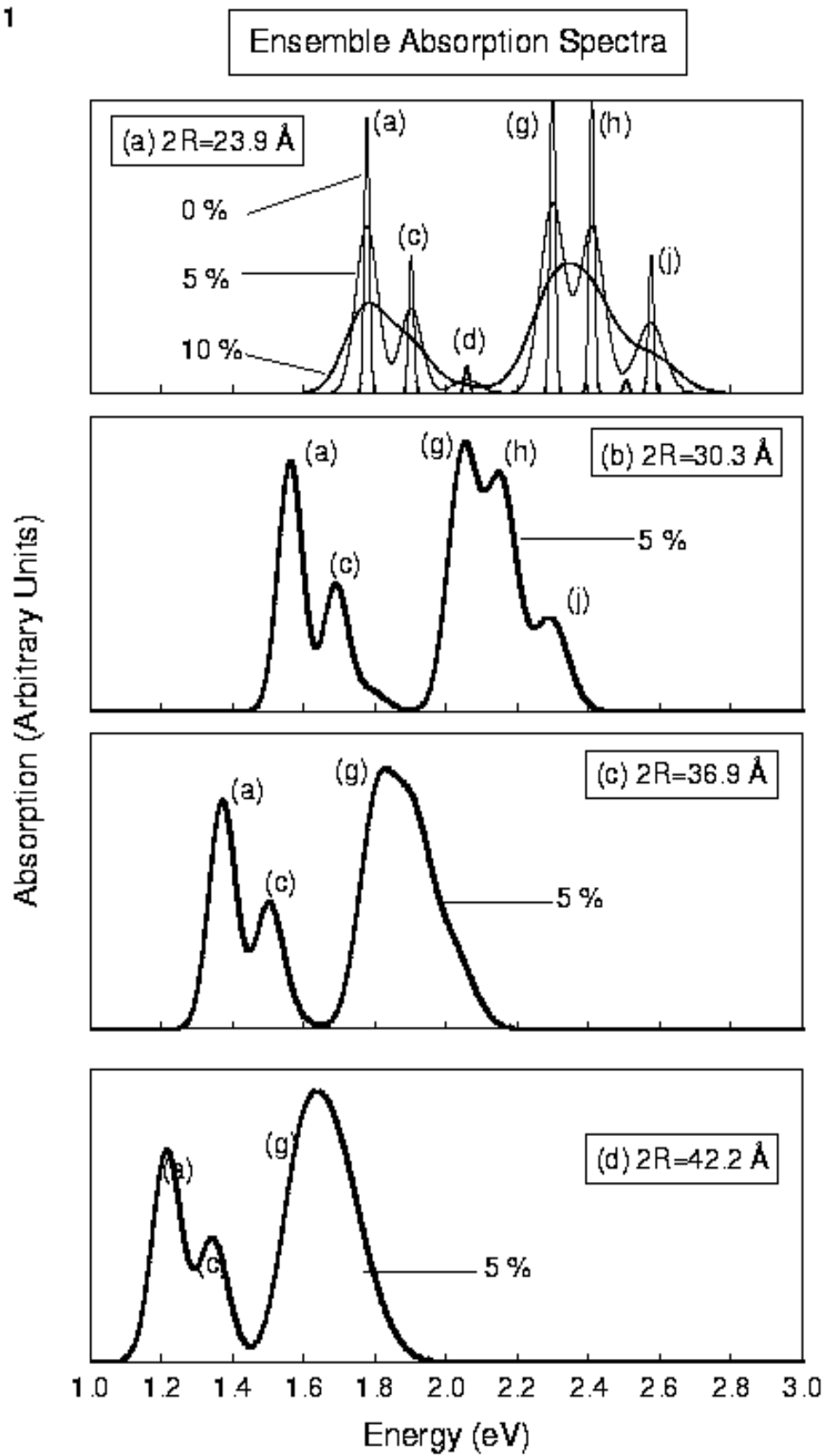
\begin{figure}
\caption{Calculated absorption spectra for ensembles of InAs quantum
dots with mean diameters of (a) 23.9, (b) 30.3, (c) 36.6 and (d)
42.2\AA.  (a) shows ensemble spectra for size distributions with 0, 5
and 10\% standard deviations.  (b), (c) and (d) show only the spectra
associated with a 5\% size distribution.  The labels (a)-(k) refer to
the classification of the states in Table~\ref{peaks_table_12}.}
\label{ensemble_spectra}
\end{figure}

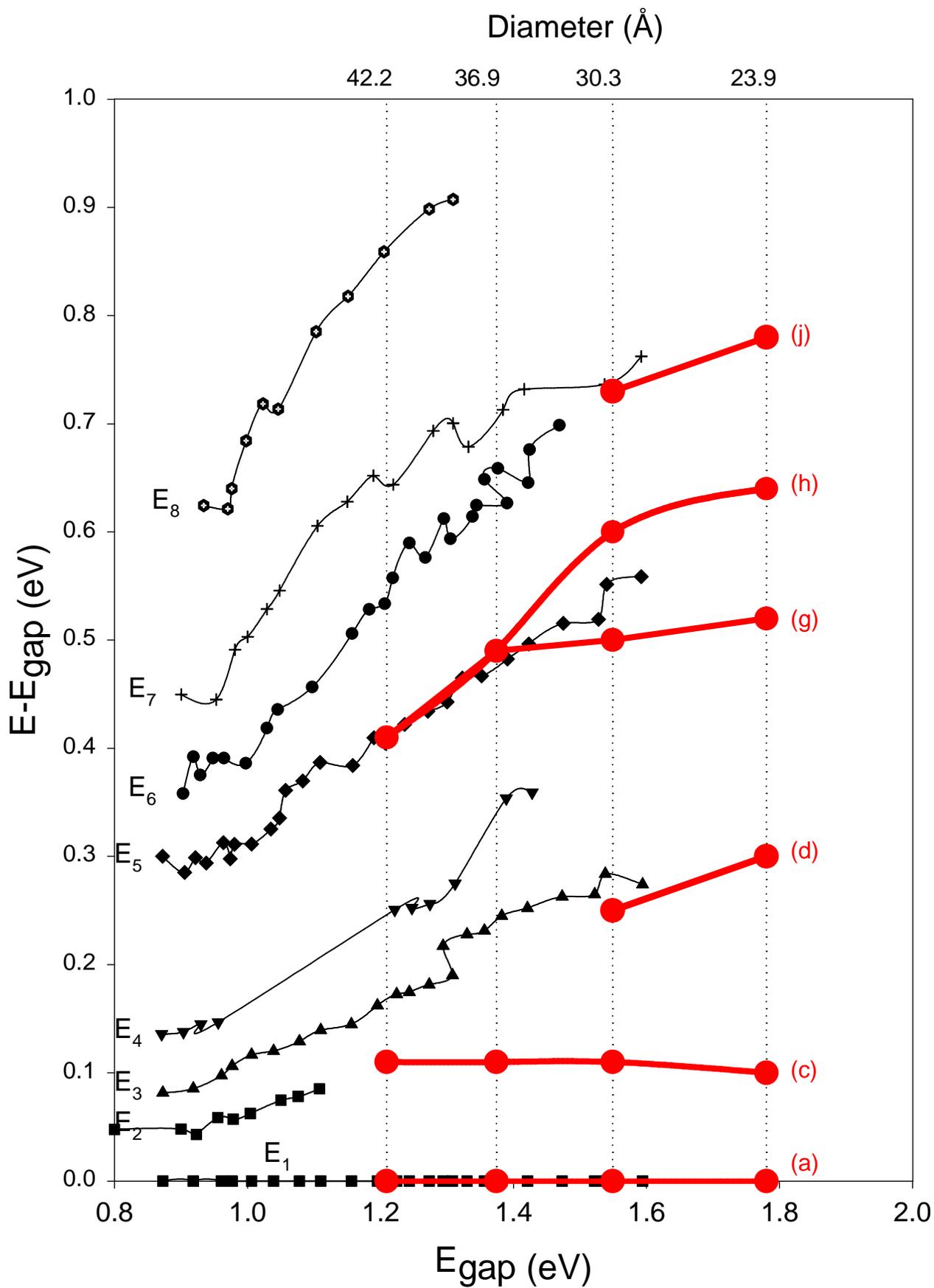
\begin{figure}
\caption{Pseudopotential calculated (circles) positions of absorption
peaks minus the band gap $vs$ band gap for InAs quantum dot ensembles
with mean diameters of 23.9, 30.3, 36.6 and 42.2\AA~and standard
deviations of 5\%.  The pseudopotential results are connected using
the analysis from section~\ref{spectra_results}.  The labels (a)-(k)
refer to the classification of the states in
Table~\ref{peaks_table_12}.  Experimental results from
Ref.\protect\onlinecite{alivisatos98} are represented by the black
lines.}
\label{ensemble_peaks}
\end{figure}

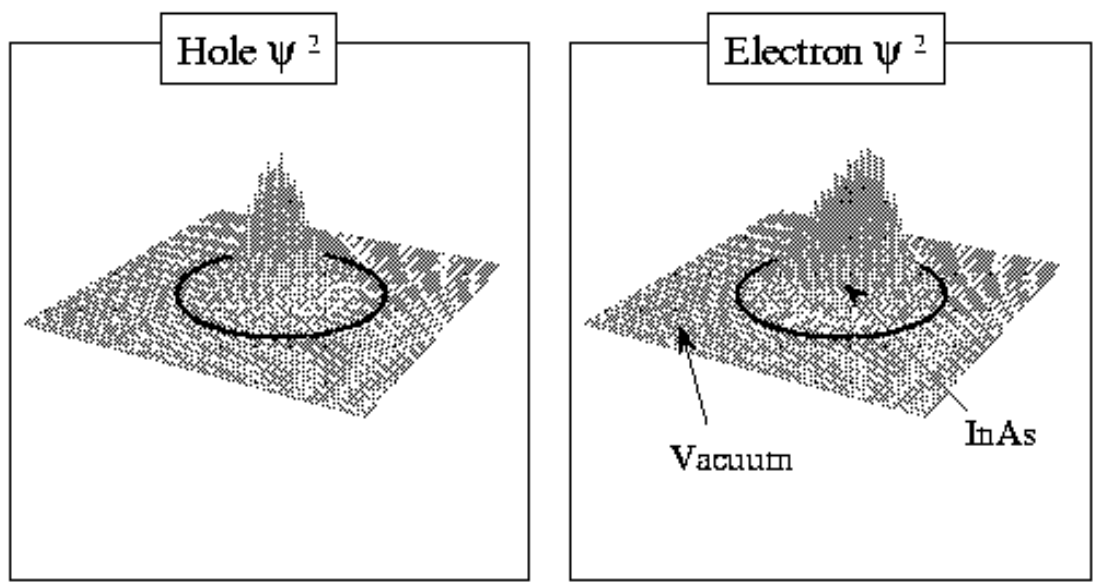
\end{document}